\documentclass{aa} % for the abstract without structuration 
\usepackage{graphicx}
\usepackage{txfonts}
\usepackage{natbib}
\bibpunct{(}{)}{;}{a}{}{,} % to follow the A&A style
\begin{document}

\title{Pulsating low-mass white dwarfs in the frame of 
new evolutionary sequences} 
\subtitle{I. Adiabatic properties}

\author{A. H. C\'orsico\inst{1,2} \and   
        L. G. Althaus\inst{1,2}}
\institute{$^{1}$ Grupo de Evoluci\'on Estelar y Pulsaciones. Facultad de 
           Ciencias Astron\'omicas y Geof\'{\i}sicas, 
           Universidad Nacional de La Plata, Paseo del Bosque s/n, 1900 
           La Plata, Argentina\\
           $^{2}$ IALP - CONICET\\
           \email{acorsico,althaus@fcaglp.unlp.edu.ar}     
           }
\date{Received ; accepted }

\abstract{Many   low-mass   white  dwarfs   with   masses
$M_*/M_{\odot}  \lesssim 0.45$,  including the  so-called extremely
low-mass white dwarfs  ($M_*/M_{\odot} \lesssim  0.20-0.25$), have
recently  been  discovered  in  the  field  of  our  Galaxy  through
dedicated   photometric  surveys.    The  subsequent   discovery  of
pulsations in  some  of  them has  opened  the unprecedented  opportunity of  
probing  the internal  structure of  these ancient stars.}   
{We  present  a  detailed
adiabatic  pulsational  study of these  stars based  on  full
evolutionary  sequences derived  from binary star evolution
computations. The  main aim  of this study  is to  provide a
detailed theoretical basis of reference for  interpreting present and
future observations  of variable low-mass white dwarfs.}{Our
pulsational analysis is based  on  a new set of He-core white-dwarf
models with  masses ranging  from  $0.1554$ to $0.4352  M_{\odot}$
derived  by computing the non-conservative evolution  of a 
binary system consisting of an initially $1 M_{\odot}$  ZAMS star 
and a $1.4 M_{\odot}$ neutron star.  We computed 
adiabatic radial ($\ell= 0$) and non-radial ($\ell= 1,
2$) $p$ and $g$ modes to assess  the dependence of the  pulsational
properties of these objects  on stellar parameters such  as the
stellar mass and the effective  temperature, as well as the effects
of element diffusion.}{We found that for white dwarf models with
masses  below $\sim 0.18 M_{\odot}$, $g$ modes  mainly probe the
core regions and $p$ modes the envelope, therefore  pulsations offer
the opportunity of constraining both the core and envelope chemical
structure of these stars via asteroseismology.  For models with $M_*
\gtrsim 0.18 M_{\odot}$, on the other hand,  $g$ modes are very
sensitive to the He/H compositional gradient and therefore can be
used as a diagnostic tool for constraining the H envelope
thickness. Because both types of objects have not only very
distinct  evolutionary histories (according to whether the
progenitor stars have experienced CNO-flashes or not), but also
have strongly different  pulsation properties, we propose to
define white dwarfs  with masses below $\sim 0.18 M_{\odot}$ as
ELM (extremely low-mass) white  dwarfs, and white dwarfs with $M_*
\gtrsim 0.18 M_{\odot}$ as  LM (low-mass)  white dwarfs.}{} 

\titlerunning{Pulsating low-mass white dwarfs}
   \maketitle
%
%________________________________________________________________

\section{Introduction}
\label{introduction}

According to the current theory of stellar evolution, most low- and
intermediate-mass \citep[$M_* \lesssim 11
  M_{\odot}$;][]{2007A&A...476..893S} stars that populate  our  Galaxy
will  end  their  lives  as   white  dwarf  (WD) stars. These old and
compact  stellar remnants have encrypted inside them a precious record  of
the evolutionary history of  the progenitor stars, providing a 
wealth of  information about the evolution of  stars,  star
formation, and  the  age  of  a variety  of  stellar populations, such
as our Galaxy and open and globular clusters  \citep{nature,
  review}. Almost 80  \% of the WDs  exhibit H-rich atmospheres;
they define the spectral class  of DA WDs.  The  mass distribution of
DA  WDs peaks at $\sim  0.59  M_{\odot}$,  and  also exhibits
high-  and  low-mass components
\citep{2007MNRAS.375.1315K,2011ApJ...730..128T,2013ApJS..204....5K}.
The  population of  low-mass WDs has  masses lower  than $0.45
M_{\odot}$ and peaks at $\sim  0.39 M_{\odot}$.  In recent years, many
low-mass  WDs with $M_*  \lesssim 0.20-0.25  M_{\odot}$ have  been detected
through the  ELM survey and the  SPY and WASP surveys
\citep[see][]{2009A&A...505..441K,2010ApJ...723.1072B,2012ApJ...744..142B,
  2011MNRAS.418.1156M,2011ApJ...727....3K,2012ApJ...751..141K};   they
are commonly referred  to as extremely low-mass  (ELM)  WDs.

Low-mass WDs are probably produced by strong mass-loss episodes at the
red  giant branch  (RGB) phase  before the  He-flash onset.  Since the
ignition of He is avoided, these  WDs are expected to harbour He cores,
in contrast to  average mass WDs, which all  probably contain C/O cores.
For  solar  metallicity progenitors,  mass-loss
episodes  must occur  in binary  systems through  mass-transfer, since
single-star  evolution is not able  to predict the  formation of these
stars in a Hubble time.  This evolutionary scenario
is confirmed by the fact that most of low-mass WDs are found in binary
systems \citep[e.g.,][]{1995MNRAS.275..828M}, and  usually  as
companions  to millisecond pulsars \citep{2005ASPC..328..357V}. In
particular, binary evolution is  the most  likely origin for  ELM WDs
\citep{1995MNRAS.275..828M}. The  evolution of  low-mass WDs   is
strongly  dependent on their stellar mass and  the occurrence of
element diffusion processes.  \citet{2001MNRAS.323..471A} and
\citet{2007MNRAS.382..779P}   have found that element diffusion  leads
to  a  dichotomy  in the  thickness  of the  H envelope, which
translates into  a dichotomy in  the age  of low-mass He-core  WDs.
Specifically,  for  stars with  $M_* \gtrsim  0.18-0.20 M_{\odot}$,
the  WD progenitor experiences  multiple diffusion-induced CNO
thermonuclear  flashes that  engulf most of  the H content  of the
envelope, and as a result,  the remnant enters the final cooling track
with  a very  thin H  envelope. As  a result,  the star  is  unable to
sustain  stable  nuclear  burning  while it cools,  and  the  evolutionary
timescale is  rather short  ($\sim 10^{7}$ yr)\footnote{Theoretical
  computations also  predict the occurrence of  H-shell flashes before
  the terminal cooling  branch is reached even if element diffusion
  processes   are  excluded, but in this case, the  H envelopes
  remain thick, with substantial H burning and long WD cooling ages
  \citep{1998A&A...339..123D,2000MNRAS.316...84S}.}.  On  the other
hand, if $M_*  \lesssim  0.18-0.20  M_{\odot}$,  the  WD  progenitor
does  not experience  H flashes  at all,  and  the remnant  enters its
terminal cooling  branch with  a thick  H envelope.  This is  thick
enough for residual  H nuclear  burning to  become the  main energy
source, which ultimately  slows  down  the  evolution,  in which  case
the  cooling timescale is  of about $\sim  10^{9}$ yr.  The  age
dichotomy has also been  suggested by observations  of the low-mass
He-core WDs  that are companions to  millisecond pulsars
\citep{2003A&A...403.1067B,2006PhDT.........7B}.

\begin{figure} 
\begin{center}
\includegraphics[clip,width=9 cm]{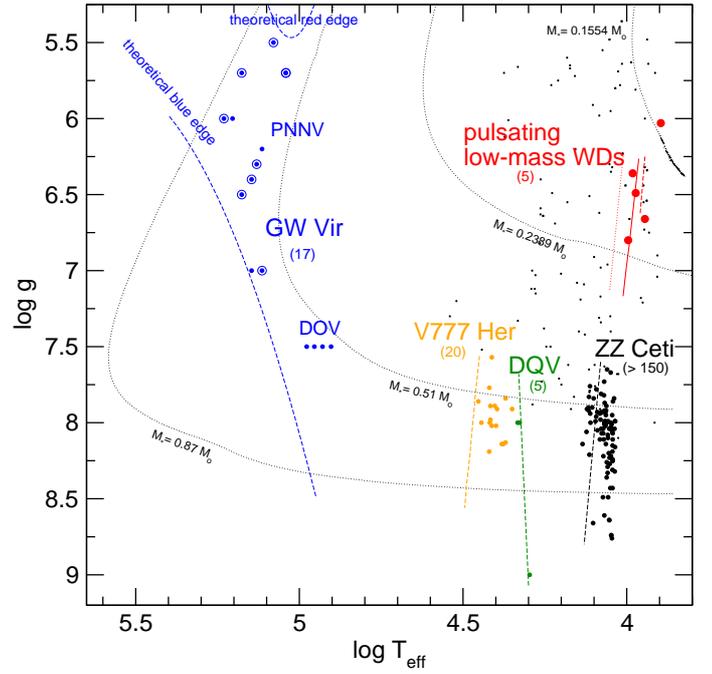} 
\caption{Location of the several  classes of pulsating WD stars in
  the $\log T_{\rm eff} - \log g$ plane, marked with dots of different
  colours.  In parenthesis we include the number of known members of
  each class.  Two post-VLTP (Very Late Thermal Pulse) evolutionary 
  tracks are  plotted   for reference.  We also   show   the  theoretical
  blue   edge  of  the instability strip for  the GW Vir stars, V777
  Her stars, the  DQV stars
  \citep[][respectively]{2006A&A...458..259C,2009JPhCS.172a2075C,
    2009A&A...506..835C}, the ZZ Ceti  stars
  \citep{2008PASP..120.1043F}, and the pulsating  low-mass  WDs.  For
  this  last  class,  we  show the  blue  edge according to
  \citet{2013MNRAS.436.3573H} (dotted  red line),
  \citet{2012A&A...547A..96C} (solid red line), and
  \citet{2010ApJ...718..441S} (dashed red line). For reference, we
  also include two evolutionary tracks of low-mass  He-core white
  dwarfs from \citet{2013A&A...557A..19A}. Small black dots 
  correspond to low-mass WDs that are non-variable or have not been 
  observed yet to assess variability.}
\label{figure-WD-all} 
\end{center}
\end{figure}

While some information  such as surface chemical composition, temperature,
and gravity  of WDs  can be inferred  from spectroscopy,  the internal
structure of these compact stars  can be unveiled  only 
 by means of asteroseismology,  an approach  based on the  comparison between
the observed  pulsation periods of  variable stars  and appropriate
theoretical models \citep{2008ARA&A..46..157W,2008PASP..120.1043F,review}.
The  first  variable WD,  HL  Tau  76,  was
serendipitously  discovered  by   \citet{1968ApJ...153..151L}.  
Since  then,  many
pulsating  WDs have  been  detected.  At  present,  there are  several
families of pulsating WDs known,  which span a wide range in effective
temperature  and gravity (Fig.  \ref{figure-WD-all}). Among  them, the
variables ZZ Ceti or DAVs (almost pure H atmospheres, $12\,500 \gtrsim
T_{\rm eff} \gtrsim 11\,000$ K)  are the most numerous ones. The other
classes comprise the DQVs (atmospheres  rich in He and C, $T_{\rm eff}
\sim 20\,000$ K),  the variables V777 Her or  DBVs (atmospheres rich in
He,  $29\,000  \gtrsim  T_{\rm  eff}  \gtrsim  22\,000$  K),  and  the
variables GW Vir (atmospheres dominated  by C, O, and He) that include
the  DOVs and  PNNVs objects  ($180\,000 \gtrsim  T_{\rm  eff} \gtrsim
65\,000$ K).  WD asteroseismology allows  us to place  constraints not
only on  global quantities such as gravity,  effective temperature, or
stellar  mass,  they  provide,  in  addition,  information  about  the
thickness of the compositional  layers, the core chemical composition,
the internal  rotation profile, the presence and  strength of magnetic
fields, the  properties of the  outer convective regions,  and several
other interesting properties  \citep[for a recent example in  the context of
ZZ Ceti stars, see][and references therein]
{2012MNRAS.420.1462R,2013ApJ...779...58R}.

The  possible  variability of  low-mass  WDs  was  first discussed  by
\citet{2010ApJ...718..441S},  who predicted  that  non-radial  pulsation
$g$ modes with long periods should be excited in WDs with masses below
$\sim 0.2 M_{\odot}$. The various attempts to find pulsating
objects of this  type were unsuccessful, however 
\citep{2012PASP..124....1S}. This
situation drastically changed with the exciting discovery of the first
pulsating  low-mass   WD,  SDSS  J184037.78+642312.3,   by  
\citet{2012ApJ...750L..28H}.  Subsequent searches  
resulted in  the detection  of two
additional  pulsating  objects,   SDSS  J111215.82+111745.0  and  SDSS
J151826.68+065813.2 \citep{2013ApJ...765..102H}, and finally another two,
SDSS  J161431.28+191219.4  and  SDSS  J222859.93+362359.6  
\citep{2013MNRAS.436.3573H},  all of them  belonging to the  
DA spectral class  and with
masses below  $0.24 M_{\odot}$. At  present, this small group  of five
stars makes up a new, separate  class of pulsating WDs.  The effective
temperatures of  these stars are found  to be between  $10\,000$ K and
$7800$ K, which means that they are  the coolest pulsating WDs known to date
(see Fig. \ref{figure-WD-all}).   Another distinctive feature of these
stars is the length of  their pulsation periods, $\Pi \gtrsim 1180$ s,
longer than the periods found in ZZ Ceti stars ($100 \lesssim \Pi
\lesssim 1200$  s).  The period  at $\Pi= 6235$  s detected in
the power  spectrum of SDSS J222859.93+362359.6 is  the longest period
ever  measured  in  a pulsating  WD  star.  On  the other  hand,  SDSS
J111215.82+111745.0 exhibits  two short periods, at $\Pi \sim 108$ s 
and $\Pi \sim 134$ s, which might be
caused by non-radial  $p$ modes or radial modes ($\ell= 0$).  
If  so,  this would  be  the  first detection  of
$p$ or radial modes  in a  pulsating WD  star.  Finally, the  
existence of  many non-variable  stars in the  region where  pulsating 
objects  are found (Fig.  \ref{figure-WD-all}) suggests  
that the  instability  strip for
low-mass WDs probably is not pure \citep{2013MNRAS.436.3573H}.  This might be a
hint that low-mass WDs that populate  the same region in the $\log T_{\rm
  eff}-\log  g$  plane   have substantially  different  internal
structures and consequently quite different evolutionary origins.

The discovery  of pulsating low-mass WDs  constitutes a unique  
opportunity to probe the  interiors of  these stars and  eventually to 
test their  formation channels by employing  the tools of 
asteroseismology. A  first step in the theoretical study of these 
variable stars has been  given by \citet{2012A&A...547A..96C},  who have
thoroughly explored the $g$-mode adiabatic pulsation properties of the
low-mass He-core  WD models with  masses in the range  $0.17-0.46 M_*$
coming from  high-metallicity progenitors ($Z= 0.03$)  and single-star
evolution computations. These authors also performed non-adiabatic
pulsation computations and  found many unstable $g$ and $p$ modes
approximately at the  correct effective temperatures and the  correct range of
the periods observed  in pulsating low-mass WDs. This  result has
later been confirmed by \citet{2013ApJ...762...57V}. 

To fully exploit the  
asteroseismological potential of this type of stars, accurate and 
realistic stellar 
models of low-mass  WDs are  crucial.  To assess the correct thermo-mechanical structure of
the WD  and the thickness of  the H envelope  left by the  pre-WD stage,  
the  complete evolutionary history  of the progenitor 
stars must  be fully accounted for. A  relevant physical ingredient  
that must be  considered during the WD  cooling phase is  
element diffusion  to consistently
account for the evolving shape of the internal chemical profiles (and,
in particular,  the chemical transition regions). Last  but not least,
stable H  burning, which is particularly  relevant in the  case of ELM
WDs  (and might  play a role  in driving  the pulsations),
must be taken into account as well.

Motivated by  the asteroseismological potential  of pulsating low-mass
WDs,  and stimulated  by  the  discovery of  the  first five  variable
objects of  this type, we report a further step  in the
theoretical study of these  stars by exploring the adiabatic pulsation
properties of  a new  set of low-mass,  He-core WD models  with masses
ranging from $0.1554$ to  $0.4352 M_{\odot}$ (including the mass range
for ELM  WDs, $M_*/M_{\odot}  \lesssim 0.20-0.25$, see  Table \ref{table1})
presented  by  \citet{2013A&A...557A..19A},  which  were  derived  by
computing the non-conservative evolution  of a binary system consisting
of an initially $1 M_{\odot}$  ZAMS star and a $1.4 M_{\odot}$ neutron
star for  various initial orbital periods. 
We here extend the study of  
\citet{2012A&A...547A..96C} in two ways.   First,  we  explore 
not  only  the non-radial  $g$-mode
pulsation  spectrum of  low-mas WD  models, but  we also 
study the non-radial  $p$ modes and  the radial  ($\ell= 0$)  pulsation 
spectrum.
Second, thanks to the availability  of five WD model sequences with
progenitors  that  have  not  suffered  from  CNO-flashes  (see  Table
\ref{table1}),  we  are now  able  to  explore  the  pulsation
properties of ELM  WDs in detail, which are characterised by very thick  H envelopes. This is at variance with the work of \citet{2012A&A...547A..96C}, 
in which only  one WD  sequence  (with mass  $M_*=  0.17 M_{\odot}$)  
corresponded to a progenitor that did not experience H flashes.  In this
paper, we examine the possible differences in the pulsation properties
of models with  stellar masses near $\sim 0.18-0.2  M_{\odot}$, which 
 could be used as a seismic tool to distinguish stars that 
have undergone CNO
flashes  in their early-cooling phase  from those  that have  not. We
additionally discuss  how our models match the  observed properties of
the known five pulsating low-mass  WD stars.  In particular, we
determine whether our models are able to account for the short periods
exhibited  by SDSS  J111215.82+111745.0 and  evaluate  the possibility
that these modes might be  $p$ modes and/or radial modes.  
We also examine the hypothesis
that one of  these stars, SDSS J222859.93+362359.6,  is  not a genuine
ELM  WD, but  instead  a He  pre-WD  star undergoing   a CNO  flash
episode. We defer to a second paper of this series a thorough
non-adiabatic exploration of our complete set of He-core WD models.

The paper is organised as  follows:  In Sect.  \ref{models} we briefly
describe  our  numerical  tools   and  the  main  ingredients  of  the
evolutionary sequences  we use to  assess the pulsation  properties of
low-mass He-core WDs.  In Sect.  \ref{pulsations} we present our pulsation results in detail.  Section  \ref{observations} is devoted to assessing
how    our    models   match    the    observations.    Finally,    in
Sect. \ref{conclusions} we summarise our main findings.

\section{Computational tools and evolutionary sequences}
\label{models}

\subsection{Evolutionary code and input physics}

\begin{figure*} 
\begin{center}
\includegraphics[clip,width=18 cm]{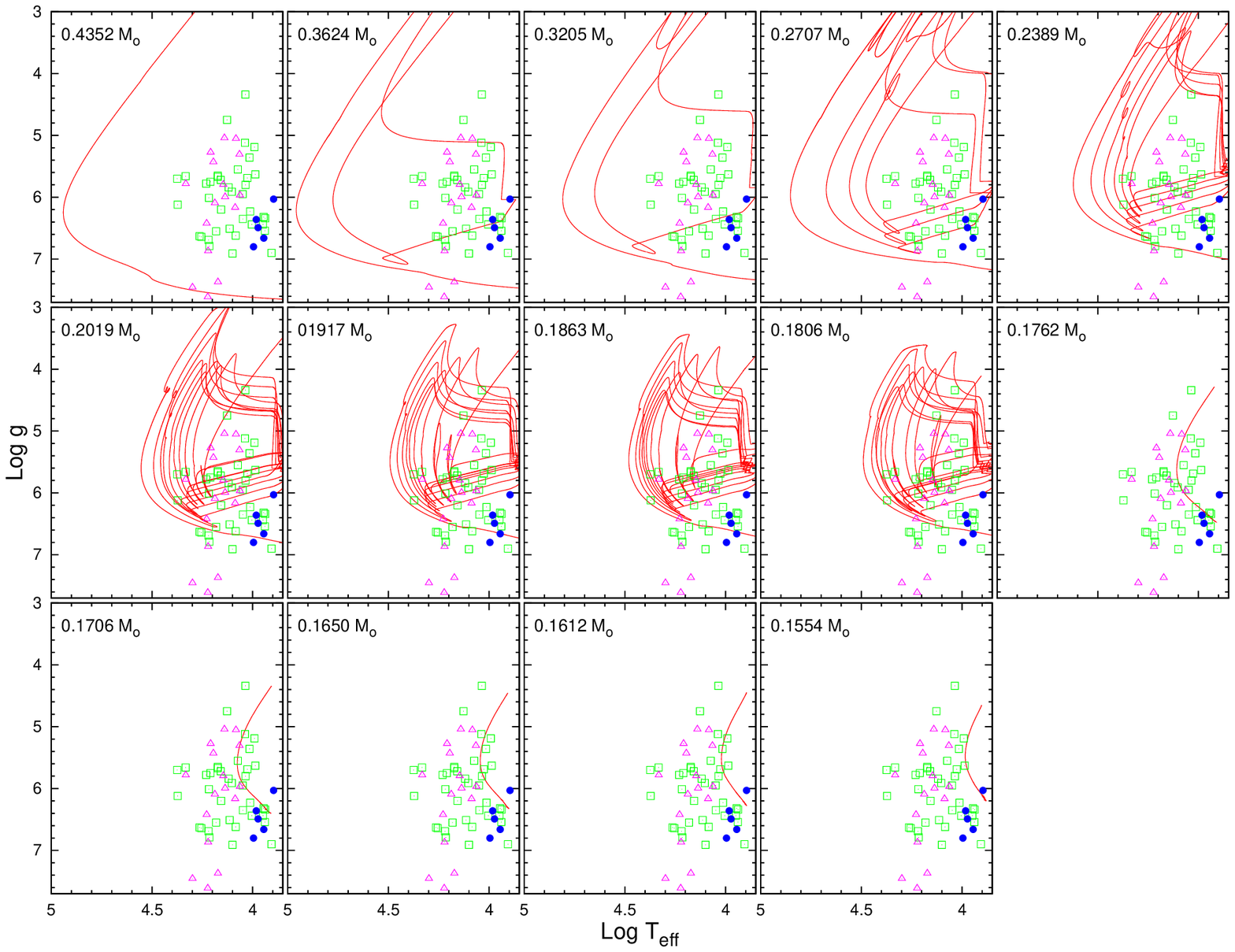} 
\caption{ $\log T_{\rm eff}- \log g$  diagrams for the He-core WD 
sequences computed in \citet{2013A&A...557A..19A}. Sequences with masses in 
the range $0.18 \lesssim M_* \lesssim 0.4$ undergo CNO flashes during 
the early-cooling phase, which leads to the complex loops in the diagram. 
Green squares and magenta triangles correspond 
to the observed post-RGB low-mass stars from \citet{2012MNRAS.424.1752S} 
and \citet{2013ApJ...769...66B}, and filled blue circles  correspond to
the five pulsating low-mass WDs detected so far 
\citep{2013MNRAS.436.3573H}. Numbers in the left upper corner 
of each panel correspond to the stellar mass at the WD stage.}
\label{figure-Tracks_Multi_II} 
\end{center}
\end{figure*} 

The  evolutionary WD models  employed  in our  pulsational analysis  were
generated with  the evolutionary code {\tt LPCODE} ,  which produces complete
and detailed  WD models that incorporate updated physical
ingredients.  While detailed information  about {\tt LPCODE} can be found in
\citet{2005A&A...435..631A,2009A&A...502..207A,2013A&A...557A..19A}  
and references therein, we list below only  the  
ingredients employed  that  are  relevant for  our  analysis  of 
low-mass, He-core WD stars.
\begin{itemize}

\item  [-] We adopted the standard  mixing length  theory (MLT)  for convection
\citep[see, e.g.,][]{2013sse..book.....K}  with the free parameter 
$\alpha = 1.6$.
With this value, the present luminosity and effective temperature
of the Sun, $\log T_{\rm eff} = 3.7641$ and $L_{\odot} = 3.842 \times 
10^{33}$ erg s$^{-1}$, at an age of 4570 Myr, are reproduced by 
{\tt LPCODE} when $Z = 0.0164$ and $X = 0.714$ are adopted, in 
agreement with the $Z/X$ value of \citet{1998SSRv...85..161G}.
A different convective efficiency that could 
change the temperature profiles  of our models would 
have no direct impact on their adiabatic pulsation properties.

\item [-] We assumed the metallicity of the progenitor  stars to be $Z =
  0.01$.

\item [-] Radiative opacities for arbitrary metallicity in the range
  from 0 to  0.1 were taken from the OPAL project  
  \citep{1996ApJ...464..943I}.
  At  low temperatures, we  used the  updated molecular  opacities with
  varying C/O ratios computed at Wichita State University 
  \citep{2005ApJ...623..585F} that were presented by 
  \citet{2009A&A...508.1343W}.

\item [-] The conductive opacities were taken from  \citet{2007ApJ...661.1094C}.

\item [-] The  equation of state during the  main-sequence evolution
  is that of OPAL for a H- and He-rich composition.

\item   [-]   Neutrino  emission   rates   for   pair,  photo,   and
  bremsstrahlung processes were taken from \citet{1996ApJS..102..411I}, 
and for plasma processes we included the treatment of 
\citet{1994ApJ...425..222H}.

\item [-] For the WD  regime we employed an updated version
  of the \citet{1979A&A....72..134M} equation of state.

\item [-] The nuclear network  takes into account 16 elements and 34
  thermonuclear  reaction  rates   for  pp-chains,  CNO  bi-cycle,  He
  burning, and C ignition.

\item [-] Time-dependent diffusion caused by gravitational settling and
  chemical  and thermal diffusion  of nuclear  species was  taken into
  account  following  the  multi-component  gas  treatment  of  
  \citet{1969fecg.book.....B}.

\item  [-]  Abundance changes  were  computed  according to  element
  diffusion, nuclear reactions,  and convective mixing.  This detailed
  treatment  of abundance  changes by  different processes  during the
  WD regime constitutes a key aspect in evaluating the
  importance of  residual nuclear burning for the  cooling of low-mass
  WDs.

\item [-]  For the  WD regime and  for effective  temperatures lower
  than $10\, 000$ K, outer boundary conditions for the evolving models
  were derived from non-grey model atmospheres \citep{2012A&A...546A.119R}.

\end{itemize}

\subsection{Pulsation codes}
\label{pulsation-codes}

We carried out a detailed adiabatic radial ($\ell= 0$) and
non-radial  $p$- and $g$-mode ($\ell= 1, 2$) pulsation analysis.    
The pulsation computations were performed with the adiabatic 
version of the pulsation code   {\tt LP-PUL} that is described in detail 
in \citet{2006A&A...454..863C}, which is
coupled to the {\tt LPCODE} evolutionary  code.   The  pulsation
code  is  based  on  a  general Newton-Raphson  technique that  solves
the full  fourth-order set  of real equations  and   boundary
conditions  that govern   linear,  adiabatic, radial, and non-radial
stellar  pulsations following the   dimensionless formulation of
\citet{1971AcA....21..289D} \citep[see][]{1989nos..book.....U}.   The
prescription we followed to  assess the run  of the Brunt-V\"ais\"al\"a
frequency ($N$) for a degenerate  environment typical of the deep
interior of a WD  is  the so-called Ledoux modified treatment
\citep{1990ApJS...72..335T}. For our exploratory stability  analysis
described in Sect. \ref{genuine},  we employed the non-adiabatic
version of the code {\tt LP-PUL} described in 
\citet{2006A&A...458..259C}.  The  code solves  the  full  sixth-order
complex system  of linearised equations and  boundary conditions as
given by  \citet{1989nos..book.....U}. The caveat  is that our
non-adiabatic  computations  rely on the frozen-convection
approximation,  in which  the  perturbation of  the convective flux
is neglected. While  this approximation is  known to give unrealistic
locations of the $g$-mode red edge of instability,  it leads to
satisfactory predictions for  the location of the blue   edge of the
ZZ Ceti (DAV) instability strip  \citep[see,
  e.g.,][]{1999ASPC..173..329B,2012A&A...539A..87V} and also for the
V777 Her (DBV) instability strip   \citep[see, for
  instance,][]{1999ApJ...516..887B,2009JPhCS.172a2075C}.

\subsection{Evolutionary sequences}

\begin{table}
\centering
\caption{Selected  properties  of our  He-core  WD  sequences
  (final cooling branch)   at  $T_{\rm eff}
  \approx 10\,000$  K: the stellar  mass, the mass  of H in  the outer
  envelope, the time
  it takes the  star to cool from $T_{\rm eff} \approx  10\, 000$ K to
  $\approx 8000$ K, and the occurrence (or not) of CNO flashes on the early 
   WD cooling branch.}
\begin{tabular}{cccc}
\hline
\hline
\noalign{\smallskip}
 $M_*/M_{\odot}$  & $M_{\rm H}/M_*\ [10^{-3}]$ & 
$\tau\ [{\rm Gyr}= 10^9 {\rm yr}]$ & H-flash\\
\hline
\noalign{\smallskip}
0.1554 & 25.4 & 3.13 &  No \\ 
0.1612 & 20.6 & 4.44 &  No \\
0.1650 & 18.7 & 5.53 &  No \\
0.1706 & 16.3 & 6.59 &  No \\
0.1762 & 14.5 & 7.56 &  No \\
\hline
\noalign{\smallskip}
0.1806 & 3.68 & 0.34 & Yes \\
0.1863 & 4.36 & 0.37 & Yes \\
0.1917 & 4.49 & 0.35 & Yes \\
0.2019 & 3.80 & 0.32 & Yes \\
0.2389 & 3.61 & 0.62 & Yes \\
0.2707 & 1.09 & 0.33 & Yes \\
0.3205 & 1.60 & 0.91 & Yes \\
0.3624 & 0.80 & 0.58 & Yes \\
0.4352 & 0.63 & 0.91 & No  \\
\noalign{\smallskip}
\hline
\hline
\end{tabular}
\label{table1}
\end{table}

To derive  realistic  configurations  for the low-mas 
He-core WDs, \citet{2013A&A...557A..19A} mimicked the binary evolution 
of progenitor stars. Since H-shell burning is the main 
source of star luminosity during most of the evolution of ELM WDs,
computating realistic initial WD structures is a
fundamental requirement, in particular for correctly assessing the H-envelope mass left by progenitor evolution
\citep[see][]{2000MNRAS.316...84S}.

\begin{figure} 
\begin{center}
\includegraphics[clip,width=9 cm]{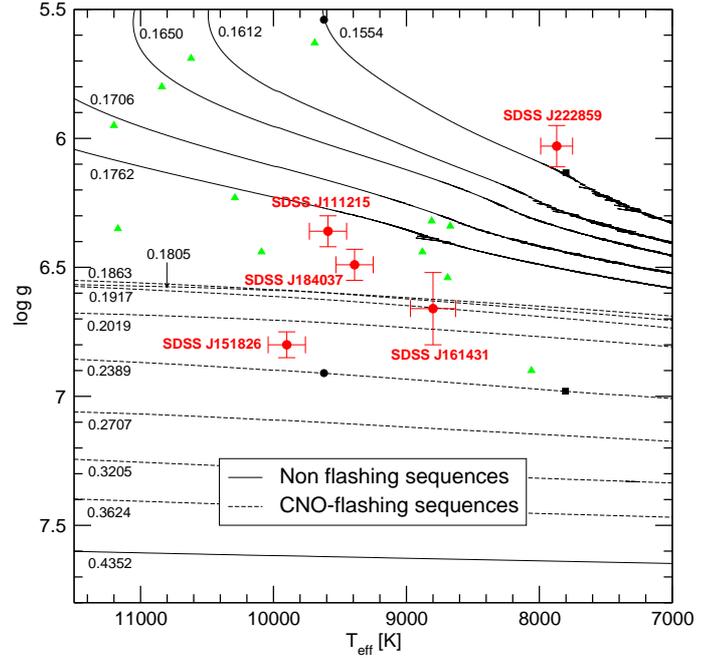} 
\caption{$T_{\rm eff}  -  \log g$  plane  showing the  low-mass
  He-core  WD evolutionary  tracks of  \citet{2013A&A...557A..19A}
  (thin black lines). Sequences with H flashes during
  the early-cooling phase are depicted with dashed lines, sequences
  without H flashes are displayed with solid 
  lines.  Numbers  correspond to the  stellar mass of  each sequence.
  The  locations of  the  five known  pulsating  low-mass WDs
  \citep{2013MNRAS.436.3573H} are marked with a small circle (red). 
  Stars not observed to vary are depicted with green triangles. Black 
  circles and squares on the evolutionary tracks of 
  $M_*= 0.1554 M_{\odot}$ and  $M_*= 0.2389 M_{\odot}$ indicate the 
  location of the template models analysed in 
  Sect. \ref{template}.}
\label{figure-HR1} 
\end{center}
\end{figure}

Binary evolution was assumed to be fully non-conservative, and
the loss of angular momentum through mass loss, gravitational-wave 
radiation, and magnetic braking was considered. All of
the He-core WD initial models were derived from evolutionary
calculations for binary systems consisting of an evolving
low-mass component of initially $1 M_{\odot}$ and a $1.4 M_{\odot}$ 
neutron star as the other component. Metallicity was assumed to be 
$Z = 0.01$. A total of 14 initial He-core WD models with stellar masses
between $0.155$ and $0.435 M_{\odot}$ were computed for initial orbital
periods at the beginning of the Roche-lobe phase in the range $0.9$
to $300$ d. While full  details about the procedure to obtain the
initial models are  provided in \citet{2013A&A...557A..19A},  
we repeat here the prescriptions used by these authors to 
obtain the  He-core WD models employed in this work.
The formalism of \citet{2000MNRAS.316...84S} is followed. If $M_1$ is the mass 
of the secondary  (mass-losing) star, $M_2$ the mass of the neutron 
star (primary), and $\dot M_1$ the mass-loss rate of the secondary, 
  the change of the total orbital angular momentum ($J$) 
of the binary system can be written as

\begin{equation}
\frac{\dot  J}{J}=  \frac{\dot J_{\rm ML}}{J}  +  \frac{\dot J_{\rm GR}}{J}  +
\frac{\dot J_{\rm MB}}{J}\, ,
\end{equation}

where   $\dot J_{\rm ML}$, $\dot J_{\rm GR}$, and  $\dot J_{\rm MB}$ 
are the angular momentum loss from the system through mass
loss, gravitational-wave radiation, and magnetic braking 
(which is relevant when the secondary has  an outer convection zone). 
To compute these quantities we follow 
\citet{2000MNRAS.316...84S} \citep[see also][]{1993MNRAS.262..164M}:

\begin{equation}
\frac{\dot J_{\rm ML}}{J} = \frac{M_2}{M_1 (M_1+M_2)} {\dot M_1}\, {\rm yr^{-1}},
\end{equation}
\begin{equation}
\frac{\dot J_{\rm GR}}{J}  = - 8.5  \times 10^{-10} 
\frac{M_1  M_2 (M_1+M_2)}{a^4}\, {\rm yr^{-1}},
\end{equation}

\begin{equation}
\frac{\dot J_{\rm MB}}{J}  = - 3  \times 10^{-7} \frac{(M_1+M_2)^2  R^4_1}{M_1 M_2
a^5}\, {\rm yr^{-1}}\, ,
\end{equation}

where $a$  is the semi-axis  of the orbit and $R_1$ the radius of the secondary.
All quantities are given in solar units.  The mass-loss rate from the secondary
is  calculated   as  in \citet{2002MNRAS.335..948C}.   Mass loss is 
considered as long as the secondary fills its Roche lobe
$r_{\rm L}$, given by

\begin{equation}
r_{\rm L} = a \frac{0.49 q^{2/3}}{0.6 q^{2/3} + \ln(1+q^{1/3})},
\label{eq.egg}
\end{equation}
where $q=M_1/M_2$  is the mass ratio. The semi-axis of the orbit is found
by integrating the equation for  the rate of change of $a$. If  mass   
lost  by  the  secondary  is
completely lost from the system, meaning that nothing of the mass lost by
the secondary is accreted by the primary, $a$ is given by 
\citep[see ][]{1993MNRAS.262..164M}

\begin{equation}
\frac{1}{2}\frac{\dot a}{a}=  \frac{\dot J}{J}  - \left(\frac{1}{M_1} -  \frac{
1}{2(M_1+M_2)}\right) \dot M_1
.\end{equation}

Mass loss is continued until the secondary star shrinks  within 
its Roche lobe.

In  Table \ref{table1}, we  provide some  main characteristics  of the
whole  set of  He-core  WD models.   The   evolution of
these  models was computed down to the range of luminosities of
cool WDs,  including the stages of multiple  thermonuclear CNO flashes
during the beginning of the cooling branch. 
The unstable 
H burning in CNO flashes occurs at the
innermost tail of the H-abundance distribution. As
the WD evolves along the cooling branch, chemical diffusion carries 
some  H inwards to hotter and C-rich layers, where it burns unstably. 
In this region, the C abundance distribution is not 
modified by  gravitational settling
\citep{2004MNRAS.347..125A}. Column 1 of Table \ref{table1}
shows the resulting final stellar masses ($M_*/M_{\odot}$). The second
column corresponds to the total  amount of H contained in the envelope
($M_{\rm H}/M_*$)  at $T_{\rm eff} \approx  10\, 000$ K (at the final 
cooling branch),  and Col. 3
displays the time spent by the models to cool from $T_{\rm eff} \approx
10\,  000$ K to  $\approx 8000$  K.  Finally,  Col. 4  indicates the
occurrence (or not) of CNO flashes on the early WD cooling branch.  We
note that, in good agreement with previous studies 
\citep{2000MNRAS.316...84S,2001MNRAS.323..471A,2007MNRAS.382..779P} 
there exists  a threshold  in the  stellar mass
value (at $\sim 0.18 M_{\odot}$), below which CNO flashes on the early
WD cooling branch are not expected to occur. Sequences 
with $M_* \lesssim 0.18
M_{\odot}$ have thicker H envelopes and much longer cooling timescales
than sequences with stellar masses  above that threshold in mass.  
In numbers, this means that the H content is about four times higher      and $\tau$ (the time to cool
from $T_{\rm eff} \approx  10\, 000$ K to $T_{\rm eff} \approx  8000$ K) is   about 22 times longer for the sequence with $M_*= 0.1762 M_{\odot}
$ than for the sequence with 
$M_*= 0.1806 M_{\odot}$ (see Table \ref{table1}). Note that in this example, 
we compare the properties of two sequences with virtually the same
stellar mass ($\Delta M_* \approx 4 \times 10^{-3} M_{\odot}$). The sequences without flashing evolve this slowly because the residual H-shell burning is the main source of surface
luminosity, even  at very  advanced stages of  evolution. It  is clear
that for  these WDs, an appropriate treatment  of progenitor evolution
is required  to correctly assess  the evolutionary timescales.
In contrast, an entirely  different behaviour is expected for sequences
that  experience  unstable H -shell  burning  on  their early  cooling
branch.  During the  final  cooling branch,  evolution  proceeds on  a
much shorter timescale  than that characterising the 
sequences with $M_* \lesssim 0.18 M_{\odot}$.  This
is because CNO flashes markedly reduce the H content of the star, with
the result  that residual nuclear  burning is much less  relevant when
the remnant reaches the final cooling  branch. 
The $\log T_{\rm eff}- \log g$ diagrams for all the He-core WD sequences 
are shown in Fig. \ref{figure-Tracks_Multi_II}, 
which is an update of Fig. 2 of \citet{2013A&A...557A..19A}. 
Sequences  that undergo CNO flashes during the
early-cooling phase exhibit complex loops in the diagram.  In
contrast, sequences  without flashes  show very simple
cooling tracks.

The mass limit  below which low-mass WDs are classified as ELM
WDs is not yet very clear  in the literature. For instance, 
\citet{2010ApJ...723.1072B,2012ApJ...744..142B,2013ApJ...769...66B}
and \citet{2013ApJ...765..102H,2013MNRAS.436.3573H} 
defined low-mass WDs with 
$M_*\lesssim 0.25 M_{\odot}$ as ELM\ WDs, while 
\citet{2011ApJ...727....3K,2012ApJ...751..141K} classified objects with $M_*\lesssim 0.20 M_{\odot  \ }$as ELM WDs. 
Here, we propose to designate as ELM WDs those low-mass WDs that do not 
experience CNO flashes in their early-cooling branch.
According to  this classification, ELM WDs 
are   low-mass WDs with masses below $\sim 0.18 M_{\odot}$ in the 
frame of our computations\footnote{Note that this theoretical 
mass threshold can 
vary according to the metallicity of the progenitor stars 
\citep{2000MNRAS.316...84S}. For instance, 
\citet{2007MNRAS.382..779P}, who assumed $Z= 0.02$ for the 
progenitor stars,
obtained a value $\sim 0.17 M_{\odot}$ for the 
mass threshold.}. This criterium is by no means
arbitrary because both types of objects not only differ in
their evolutionary history, but also show markedly different
pulsation properties. 

We conclude  this section by  showing in Fig. \ref{figure-HR1}  the $\log
T_{\rm eff}- \log g$ plane  in the region where pulsating low-mass WDs
are found, along  with the He-core low-mass WD  evolutionary tracks of
\citet{2013A&A...557A..19A}.  Note that there is a
gap between the two sets of tracks, corresponding to a stellar mass of
$\sim 0.18  M_{\odot}$. Interestingly, there are  two pulsating stars,
SDSS J184037.78$+$642312.3  and SDSS J111215.82$+$111745.0,  which are
located  precisely  in that  transition  region  between  ELM WDs  and
low-mass  WDs,  according  to  our definition.  They  are,  therefore,
very interesting targets for asteroseismology.

\section{Pulsation results}
\label{pulsations}

\begin{figure*} 
\begin{center}
\includegraphics[clip,width=17 cm]{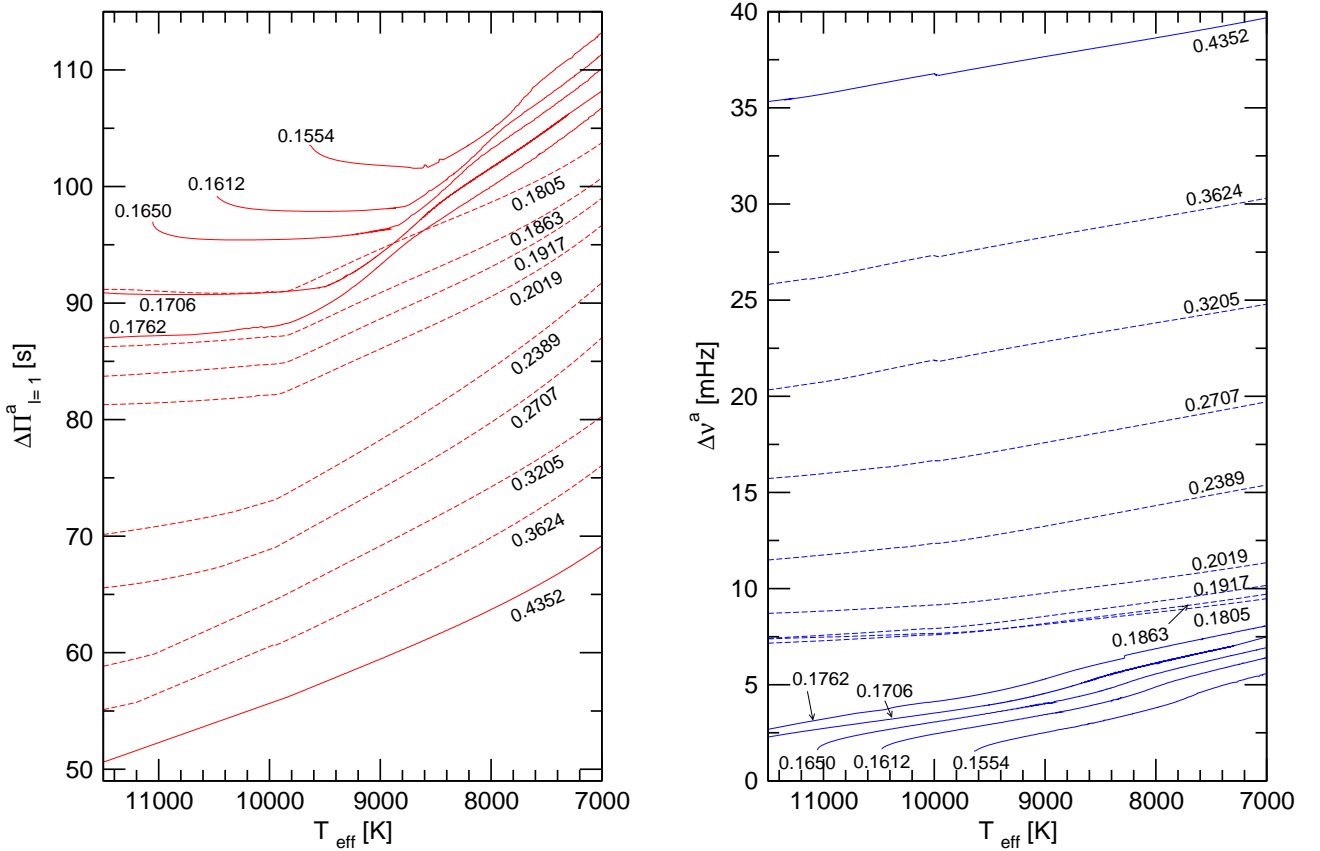} 
\caption{Dipole ($\ell= 1$)  asymptotic  period spacing 
of $g$ modes (left panel)  
and the asymptotic frequency spacing of $p$ modes (right panel)
in  terms  of the  effective temperature for all of 
our low-mass He-core evolutionary sequences. Dashed lines correspond to
sequences with CNO flashes during the early-cooling 
phase, solid lines sequences without H flashes.}
\label{figure-asintoticos} 
\end{center}
\end{figure*}

In this section we present the results of our detailed adiabatic survey of 
radial modes ($\ell= 0$) and non-radial $p$ and $g$ modes ($\ell= 1, 2$) 
for the complete set of low-mass, He-core WD sequences of 
\citet{2013A&A...557A..19A}. We cover a wide range
of effective temperatures ($12\,000 \gtrsim T_{\rm eff} \gtrsim 7\,000$ K)
and stellar masses ($0.155 \lesssim M_*/M_{\odot} \lesssim 0.435$). The set 
of the computed modes covers a very wide range of periods,  
embracing all the periodicities detected in pulsating low-mass 
WDs up to now. Specifically, the lower limit of the computed periods
is between $\Pi_{\rm min} \sim 0.5$ s (corresponding to 
$M_*= 0.4352 M_{\odot}$) and 
$\Pi_{\rm min} \sim 10$ s (corresponding to $M_*= 0.1554 M_{\odot}$). 
They are associated with high-order radial modes and $p$ modes, 
and were computed to account 
for the shortest periods detected in 
SDSS J111215.82$+$111745.0 ($\Pi \sim 108-134$ s). On the 
other hand, the upper limit 
of the computed periods (associated with high-order $g$ modes) 
is $\Pi_{\rm max} \sim 7000$ s, to account for the longest
periods detected in SDSS J222859.93+362359.6 ($\Pi \sim 6235$ s).

In the next section we discuss our pulsational results 
by showing the predictions of the asymptotic theory of stellar
pulsations.

\subsection{Asymptotic period spacing ($\Delta \Pi^{\rm a}$) and 
asymptotic frequency spacing ($\Delta \nu^{\rm a}$)}
\label{asymptotic}

For  $g$ modes  with  high   radial  order  $k$  (long  periods),  the
separation  of consecutive  periods ($|\Delta  k|= 1$)  becomes nearly
constant  at a  value  given  by the  asymptotic  theory of  non-radial
stellar  pulsations.   Specifically,  the  asymptotic  period  spacing
\citep{1990ApJS...72..335T} is given by

\begin{equation} 
\Delta \Pi_{\ell}^{\rm a}= \Pi_0 / \sqrt{\ell(\ell+1)},  
\label{aps}
\end{equation}

\noindent where

\begin{equation}
\label{asympeq}
\Pi_0= 2 \pi^2 \left[\int_{r_1}^{r_2} \frac{N}{r} dr\right]^{-1}.
\label{p0}
\end{equation}

\noindent The squared Brunt-V\"ais\"al\"a frequency ($N$, one of the 
critical frequencies of non-radial stellar pulsations) is computed as 

\begin{equation}
\label{bvf}
N^2= \frac{g^2 \rho}{P}\frac{\chi_{\rm T}}{\chi_{\rho}}
\left[\nabla_{\rm ad}- \nabla + B\right],
\label{bv}
\end{equation}

\noindent where the compressibilities are defined as

\begin{equation}
\chi_{\rho}= \left(\frac{d\ln P}{d\ln \rho}\right)_{{\rm T}, \{\rm X_i\}}\ \ \
\chi_{\rm T}= \left(\frac{d\ln P}{d\ln T}\right)_{\rho, \{\rm X_i\}}.
\end{equation}

\noindent The Ledoux term $B$ is computed as \citep{1990ApJS...72..335T}

\begin{equation}
\label{B}
B= -\frac{1}{\chi_{\rm T}} \sum_1^{M-1} \chi_{\rm X_i} \frac{d\ln X_i}{d\ln P}, 
\label{BLedoux}
\end{equation}

\noindent where
\begin{equation}
\chi_{\rm X_i}= \left(\frac{d\ln P}{d\ln X_i}\right)_{\rho, {\rm T}, 
\{\rm X_{j \neq i}\}}.
\end{equation}

\noindent The  expression in Eq.  (\ref{aps}) is  rigorously valid for
chemically  homogeneous  stars.   In   this  equation  (see  also  Eq.
\ref{asympeq}), the dependence of $\Delta \Pi_{\ell}^{\rm a}$ 
on the Brunt-V\"ais\"al\"a frequency is
such that the asymptotic period spacing is larger when the mass and/or
effective temperature of the model is lower.  This trend is clearly visible in the left panel of Fig. \ref{figure-asintoticos}, in which we depict
the evolution of  the asymptotic period spacing of $g$ modes 
for  all the sequences
considered in this work.  The higher values of $\Delta \Pi_{\ell}^{\rm
  a}$ for lower  $M_*$ comes from the dependence  $N \propto g$, where
$g$ is the  local gravity ($g\propto M_*/R_*^2$).  On  the other hand,
the  higher values of  $\Delta \Pi_{\ell}^{\rm  a}$ for  lower $T_{\rm
  eff}$ result  from the dependence  $N \propto \sqrt  \chi_{T}$, with
$\chi_{T}  \rightarrow 0$  for increasing  electronic degeneracy  
($T \rightarrow 0$). The abrupt change in the slope of the curves 
representing $\Delta \Pi_{\ell}^{\rm  a}$ at certain effective temperatures 
(that decrease for decreasing stellar mass) is due to the appearance of 
surface convection in the models, which induces a lower value of the 
integral in Eq. (\ref{p0}) and the consequent increase of 
$\Delta \Pi_{\ell}^{\rm  a}$.

\begin{figure} 
\begin{center}
\includegraphics[clip,width=9 cm]{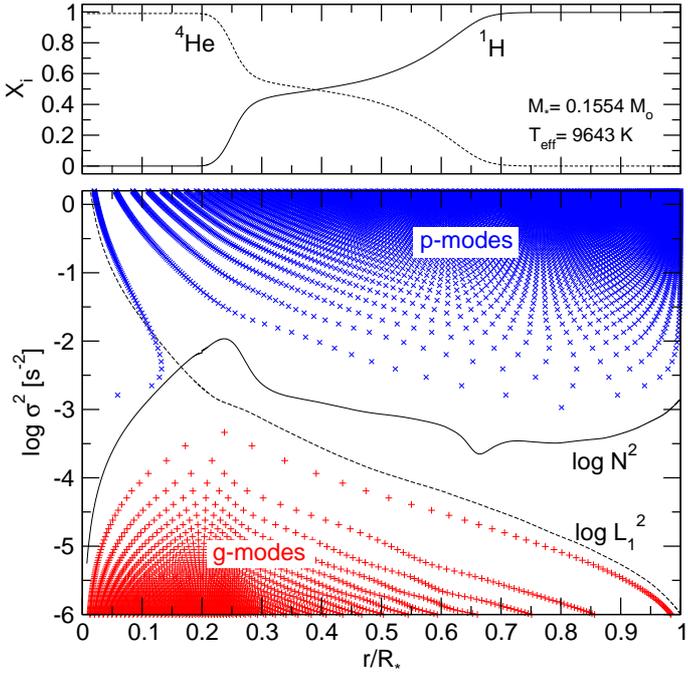} 
\caption{Internal chemical profiles of  He and H (upper panel) and
  the propagation diagram  ---the run of the logarithm  of the squared
  critical frequencies ($N$, $L_{\ell}$)---  (lower panel) 
  corresponding to  the ELM WD template  model  of $M_*=  0.1554  M_{\odot}$  
  and  $T_{\rm eff}  \approx
  9\,600$ K.  Plus symbols (in red) correspond to the
  spatial  location  of  the  nodes  of the  radial  eigenfunction  of
  dipole ($\ell= 1$) $g$ modes, x symbols (in blue) represent the
  location  of  the  nodes of dipole $p$ modes.}
\label{figure-xfbvl-10000-0.1554} 
\end{center}
\end{figure}

The strong dependence  of the period spacing on  $M_*$ shown
in Fig. \ref{figure-asintoticos}  might be  used, in principle,  to 
infer  the stellar mass  of  pulsating  low-mass   WDs,  provided  
that  enough consecutive pulsation periods of $g$ modes were detected.  
However, this prospect is severely complicated because the period
spacing of pulsating WDs also depends on the thickness of the
outer H envelope \citep{1990ApJS...72..335T},  $\Delta \Pi_{\ell}^{\rm
  a}$ being larger  for thinner envelopes.  This is  particularly true in
the context of  low-mass He-core WDs, in  which ELM WDs models,
because of the absence of CNO flashes, harbour H  envelopes that are  
several times  thicker than more   massive  models.    For the  value   of  $\Delta
\Pi_{\ell}^{\rm a}$, and for a fixed $T_{\rm eff}$, therefore,
a model with a low
mass and a  thick H envelope can readily mimic  a more massive model
with  a thinner H envelope. A clear demonstration of this ambiguity can be
found in Fig. \ref{figure-asintoticos}, which shows a notorious degeneracy 
of the asymptotic period spacing. This figure shows for instance 
that for $T_{\rm eff} \gtrsim 8500$ K, ELM WD models with 
masses $M_*= 0.1762 M_{\odot}$ and $M_*= 0.1706 M_{\odot}$ have lower
values of $\Delta \Pi_{\ell}^{\rm a}$ than the more massive models with 
$M_*= 0.1805 M_{\odot}$, which is caused by the much thicker
H envelopes of the former models. If  for a 
given pulsating star a rich  spectrum of
observed  periods were available,  this degeneracy  might  be  broken by
including  additional  information of  the  mode-trapping  properties,
which   yield   clues   about    the thickness of the   H   envelope.

The  asymptotic  frequency  spacing  of  $p$ modes ($k \gg 1$), 
on the other hand, is given by \citep{1989nos..book.....U}
\begin{equation} 
\Delta \nu^{\rm a} = \left[ 2 \int_0^R \frac{dr}{c_{\rm s}}\right]^{-1},
\label{afs}
\end{equation}

\noindent where $c_{\rm s}$ is the local adiabatic sound speed, 
defined as $c_{\rm s}^2 =\Gamma_1 P / \rho$. The asymptotic frequency 
spacing is related to the Lamb frequency (the other critical frequency
of non-radial stellar pulsations) through $c_{\rm s}$  by means
of 
\begin{equation} 
L_{\ell}^2 = \ell (\ell+1) \frac{c_{\rm s}^2}{r^2}
\label{lamb}
.\end{equation}

In the right-hand panel of Fig. \ref{figure-asintoticos} we display
the evolution of  the asymptotic frequency spacing 
(in units of mHz $\equiv 10^{-3}$ Hz) for  all the sequences
considered in this work. As can be seen, $\Delta \nu^{\rm a}$
is larger for higher stellar masses and lower effective 
temperatures. This behaviour can be understood by realising that
for higher $M_*$ and lower $T_{\rm eff}$, the values
of the sound speed of the models  globally increases, as a result of which
the value of the integral in Eq. (\ref{afs}) is lower, and consequently, 
$\Delta \nu^{\rm a}$ increases.

\begin{figure} 
\begin{center}
\includegraphics[clip,width=9 cm]{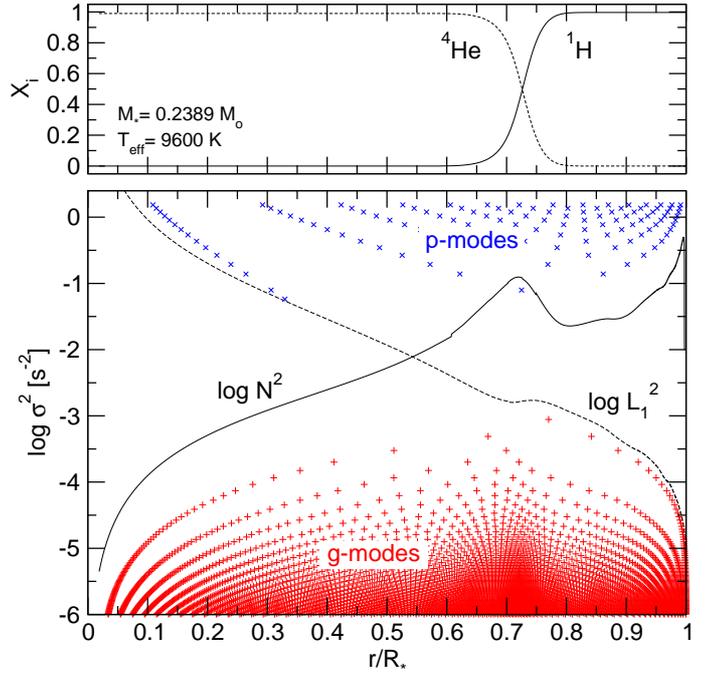} 
\caption{Same as Fig. \ref{figure-xfbvl-10000-0.1554}, but for the 
LM  WD template model of $M_*=  0.2389  M_{\odot}$. }
\label{figure-xfbvl-10000-0.2389} 
\end{center}
\end{figure}

In contrast to what occurs for the 
$\Delta \Pi_{\ell}^{\rm a}$ for $g$ modes, the asymptotic frequency spacing for $p$ modes is not
degenerate for sequences with CNO flashes ($M_* \gtrsim 0.18 M_{\odot}$)
and those without ($M_* \lesssim 0.18 M_{\odot}$, ELM WDs).
This is because $\Delta \nu_{\ell}^{\rm a}$ is rather insensitive
to the thickness of the H envelope. While an in-depth discussion 
of $p$ modes in low-mass WDs is only of academic interest as
yet, we can envisage that if these modes were confirmed in future observations, the 
eventual measurement of the mean frequency spacing for a real star 
might help in constraining  its stellar mass. 

\subsection{Template models at the edges of the instability strip}
\label{template}

\begin{figure} 
\begin{center}
\includegraphics[clip,width=9 cm]{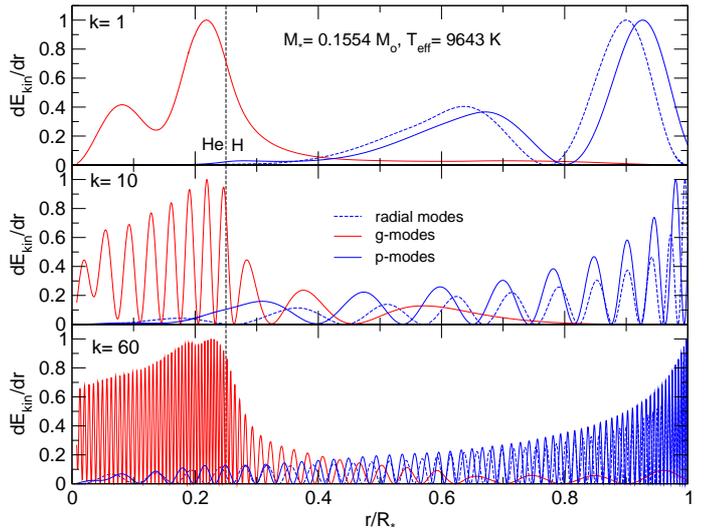} 
\caption{Run  of the density of  kinetic energy
  $dE_{\rm ekin}/dr$ (normalised to 1) for radial modes (dashed
blue) and dipole 
  $g$ (red) and $p$ modes 
  (solid blue curves) 
  with $k= 1$ (upper panel), $k=
  10$ (middle panel),  and $k= 60$ (lower panel),  corresponding to the
  ELM template model with $M_*=  0.1554 M_{\odot}$. The vertical dashed line
  marks the  location  of the  He/H
  chemical  transition  region.}
\label{figure-dens-ekin-10000-0.1554} 
\end{center}
\end{figure}

To illustrate the adiabatic pulsation properties of 
our huge set of low-mass He-core 
WD models, which comprises more than $10000$ stellar structures, we focused on 
some selected, representative models. Since the instability domain 
of the known pulsating 
low-mass WDs in the $\log T_{\rm eff}-\log g$ diagram is comprised between
$T_{\rm eff} \sim 10\,000$ K (the empirical blue edge) and 
$T_{\rm eff} \sim 7800$ K (the empirical red edge),
we considered template models at both boundaries.  

We first analyse two template models at $T_{\rm eff} \sim 9600$ K
with stellar mass  $M_*= 0.1554 M_{\odot}$ (ELM template model)
and $M_*= 0.2389 M_{\odot}$ (LM template model). The ELM template model
is representative of structures without CNO flashes in the past
evolution ($M_* \lesssim 0.18 M_{\odot}$), while the LM template model 
characterises the objects with H 
flashes ($M_* \gtrsim 0.18 M_{\odot}$). The location of these template 
models in the $\log T_{\rm eff}-\log g$ diagram is indicated in 
Fig. \ref{figure-HR1} with black circles. 
Our models have a He core surrounded by a
H outer  envelope.  In  between, there is  a smooth  transition region
shaped  by  the  action   of  microscopic  diffusion. In the  
upper panels  of Figs. \ref{figure-xfbvl-10000-0.1554}
and \ref{figure-xfbvl-10000-0.2389}  we display the internal chemical
profiles for  He and H corresponding  to the two template  
models. These template models are different in two important  
ways.  To begin with,  
the  ELM model  has  a H  envelope that  is about seven$ $ times thicker than the LM model. As mentioned, 
this  is the result  of the very  different evolutionary
history  of  the  progenitor   stars.   Second, the 
He/H transition region is markedly wider for
the ELM  model than  for the LM  one. In particular, the H profile 
for the ELM model is  characterised   by  a   diffusion-shaped
double-layered chemical structure, which consists of a pure H envelope
 on top of an  intermediate remnant  shell rich in  H and He. We
return to this in Sect. \ref{diffusion}.

\begin{figure} 
\begin{center}
\includegraphics[clip,width=9 cm]{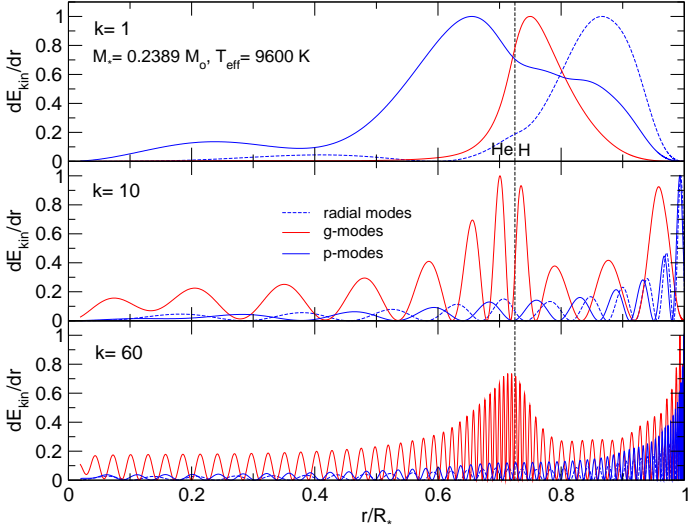} 
\caption{Same as Fig. \ref{figure-dens-ekin-10000-0.1554}, but for
 the LM WD template model with $M_*=  0.2389 M_{\odot}$.}
\label{figure-dens-ekin-10000-0.2389} 
\end{center}
\end{figure}

The chemical transition regions leave notorious  signatures in the
run of the squared critical frequencies, in particular in $N$. 
This is clearly displayed
in the propagation diagrams {\it  } 
\citep{1980tsp..book.....C,1989nos..book.....U}  of the 
lower panels in Figs.   \ref{figure-xfbvl-10000-0.1554} 
and \ref{figure-xfbvl-10000-0.2389} that correspond to the template models.  
$g$ modes propagate in the regions where
$\sigma^2  <  N^2,  L_{\ell}^2$,  and $p$ modes in the regions where
$\sigma^2  >  N^2,  L_{\ell}^2$. Here, $\sigma$  is  the  oscillation
frequency.  
Note the  very different 
shape of $N^2$  for both models. In particular, one of the  
strongest differences is the location of the bump at the He/H  
transition region. Indeed, this bump is located at much deeper 
regions for the ELM model
($r/R_* \sim 0.2$) than for the LM model ($r/R_* \sim 0.7$). 
From the local maximum  at
$r/R_* \sim 0.22$ (which coincides with the He/H transition  region)   
outwards, $N^2$  also decreases for the ELM model until it reaches 
a minimum at $r/R_* \sim 0.8$.  
After that minimum,
$N^2$ reaches higher values ​​in the surface
layers.  This  is in  contrast with the  LM template model,  
in which  the squared  Brunt-V\"ais\"al\"a  frequency exhibits
higher values at the outer layers than at the core, which resembles 
the situation in the C/O
core DAV WD models \citep[see Fig. 3 of][]{2012MNRAS.420.1462R}.  This
notoriously different shape  of the run of $N^2$  has strong consequences for
the propagation  properties of  eigenmodes.  Specifically, and  because of
the particular shape of $N^2$, which is larger in the core than in the
envelope \citep[see also Fig. 3 of][]{2012A&A...547A..96C},  
the resonant  cavity of  
$g$ modes for the  ELM model  is circumscribed  to the  core regions
($r/R_*  \lesssim  0.25$) and the opposite holds for $p$ modes,  
whereas  for  the  LM model  the
propagation region for both $g$ and $p$ modes extends roughly along 
the whole model.  To summarise, \emph{g modes in ELM WD models 
($M_* \lesssim 0.18 M_{\odot}$) mainly probe the core regions 
and $p$ modes the envelope. This means that we have the opportunity to 
constrain both the core and envelope chemical structure of these stars
via asteroseismology}. \citet{2010ApJ...718..441S} were the first to notice 
this important characteristic.

\begin{figure} 
\begin{center}
\includegraphics[clip,width=9 cm]{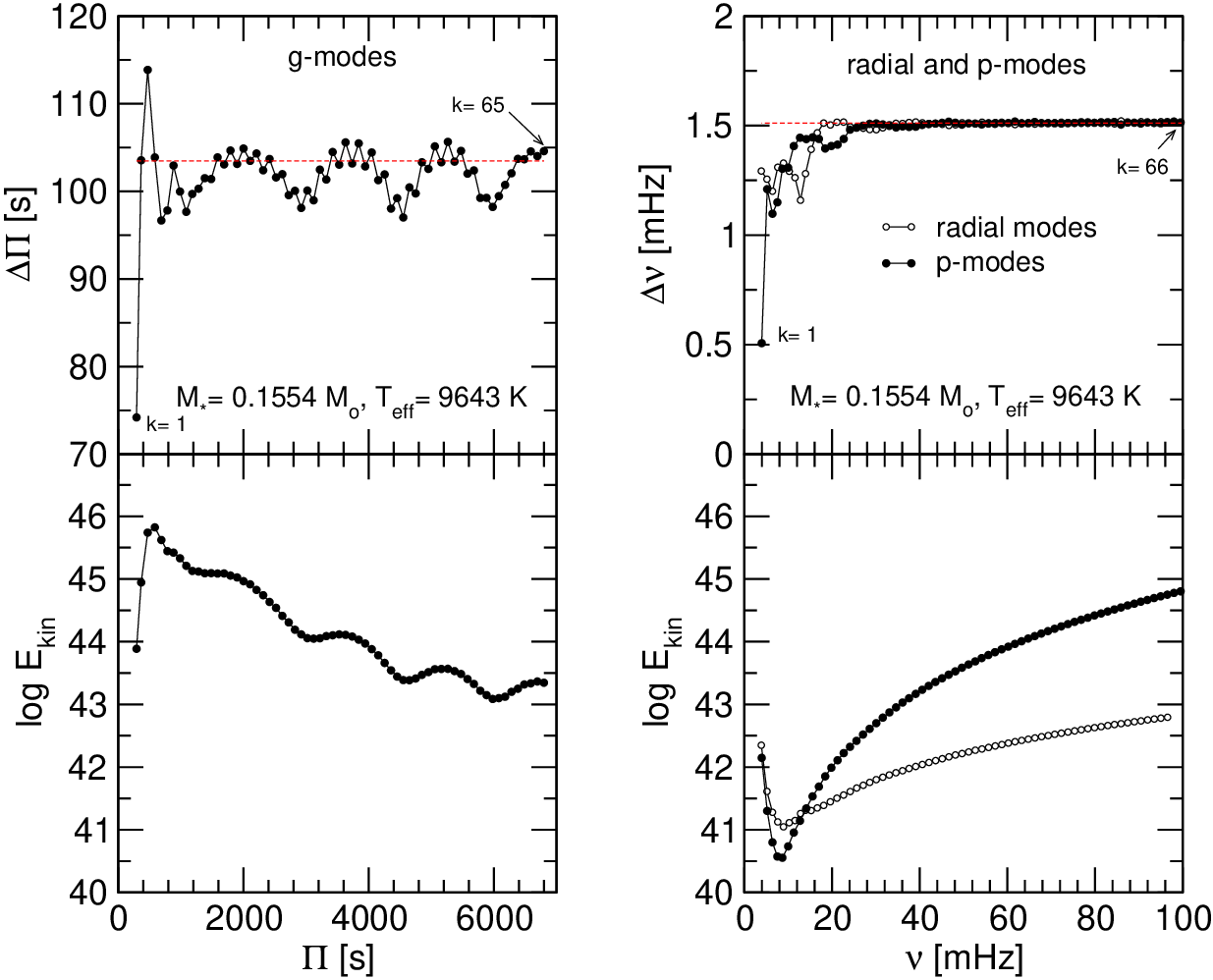} 
\caption{ $\ell= 1$ forward period-spacing  of $g$ modes versus 
   periods  
  (upper left-hand panel), and 
  the forward frequency-spacing of radial modes (hollow small circles) 
 and $p$ modes (filled small circles) versus 
  frequencies 
  (upper right-hand panel) and the associated oscillation
  kinetic energy distributions (lower panels) for 
  the ELM WD template model
  with $M_*= 0.1554  M_{\odot}$ and $T_{\rm eff}\sim  9\,600$ K.
  The  red horizontal  lines  in  the upper  panels  correspond to  the
  asymptotic period-spacing (left), computed with Eq. (\ref{aps}), 
  and the asymptotic frequency-spacing (right), computed with 
  Eq. (\ref{afs}).}
\label{figure-delp-ekin-10000-0.1554} 
\end{center}
\end{figure}

The property described above is vividly illustrated in 
Fig. \ref{figure-dens-ekin-10000-0.1554}, in which we display
the density of the oscillation kinetic energy $dE_{\rm kin}/dr$,
\citep[see Appendix A of][for its definition]{2006A&A...454..863C} 
for  radial modes, $p$ and $g$ modes with 
radial order $k= 1, 10,$ and $60$ for the same ELM template model 
as was analysed in Fig. \ref{figure-xfbvl-10000-0.1554}. 
It  is apparent  
that most  of the kinetic energy  of $g$ modes  
is confined  to the  regions below  the He/H  interface, 
meaning that  most of the spatial oscillations are located in 
the region with $r/R_* \lesssim 0.25$.  In contrast,
the kinetic energy of $p$ modes and radial modes is spread 
throughout the star, but concentrated more towards the 
surface regions and almost absent 
from the core.

In Fig. \ref{figure-dens-ekin-10000-0.2389} we show the situation 
for the LM WD template model. At variance with the ELM model,
there is no such a clear distinction in the behaviour of $g$ modes 
with respect to $p$  and radial modes.  Indeed, the
kinetic energy of $g$ modes is spread throughout the model, although
it is particularly concentrated in the region of the He/H interface and
also at the surface.  In this sense, low-mass He-core WDs with 
$M_* \gtrsim 0.18 M_{\odot}$ behave in a qualitative similar way  
as their massive cousins,
the C/O-core DAV WD stars.  $p$ and radial modes
are insensitive to the presence of the chemical interface because their
kinetic energy is enhanced at the stellar surface. In summary, it is
apparent that for LM WD models, \emph{g modes are very sensitive to
the He/H compositional gradient and  therfore can be a diagnostic
tool to constrain the H envelope thickness of low-mass WD models
with $M_* \gtrsim 0.18 M_{\odot}$}.

We now examine the mode-trapping  properties of our template models.
Mode trapping of $g$ modes is a well-studied mechanical resonance that
acts in WD stars through chemical composition gradients. 
This means that one  or more narrow regions  
in which the abundances of  nuclear species (and  consequently, the average  
molecular weight $\mu$) vary spatially modify the character 
of  the resonant cavity   in  which   modes   should  propagate   
as  standing   waves. Specifically, chemical transition regions  
act like reflecting walls that partially trap  certain modes, 
forcing  them to oscillate  with larger amplitudes in specific 
regions  and with smaller amplitudes outside of those regions 
\citep[see, for details,][]{1992ApJS...80..369B,
1993ApJ...406..661B,2002A&A...387..531C}.
Mode trapping translates into local maxima and minima in  
$E_{\rm kin}$,  which are usually
associated with  modes that are  partially confined to  the core regions  and 
modes that are partially trapped in the envelope. Unfortunately, 
the kinetic oscillation energy is hard to estimate 
from observations alone. A more important signature, which in principle 
can be employed as an observational diagnostic of mode trapping ---provided 
that a series of periods with the same $\ell$ and 
consecutive radial order $k$ is detected--- is the strong departure 
from uniform period spacing, $\Delta
\Pi$  ($\equiv  \Pi_{k+1}-\Pi_k$), when  plotted in  terms  of  the
pulsation period $\Pi_k$. For stellar  models
characterised by  a single  chemical interface, like  those  we 
consider here, local minima in $\Delta \Pi_k$ usually correspond to
modes trapped  in the H envelope, whereas  local maxima  in $\Delta
\Pi_k$ are associated with modes trapped in the core region.

\begin{figure} 
\begin{center}
\includegraphics[clip,width=9 cm]{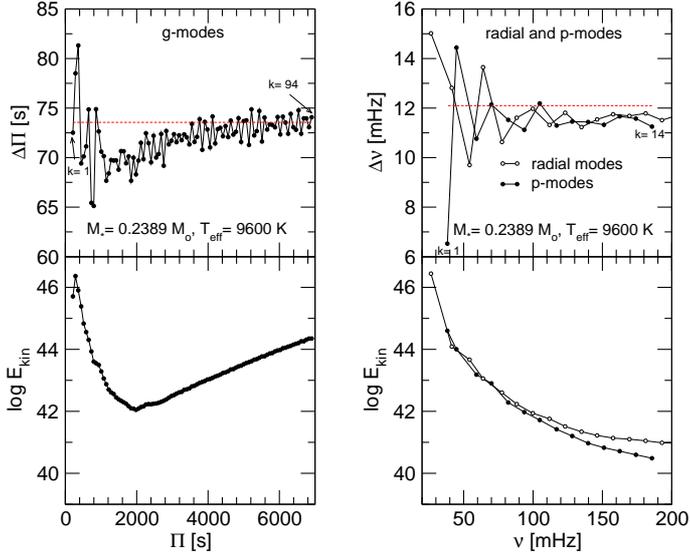} 
\caption{Same as Fig. \ref{figure-delp-ekin-10000-0.1554}, but for 
the LM WD template model with $M_*= 0.2389  M_{\odot}$ and 
$T_{\rm eff}\sim  9\,600$ K.}
\label{figure-delp-ekin-10000-0.2389} 
\end{center}
\end{figure}

In   the   upper left-hand  panel   of   Fig.    
\ref{figure-delp-ekin-10000-0.1554}   we  show the dipole forward  
period-spacing of $g$ modes versus periods for the ELM template model
with $M_*= 0.1554  M_{\odot}$ and $T_{\rm eff}\sim  9\,600$ K. 
The upper right-hand
panel corresponds to  the  forward  frequency-spacing, $\Delta
\nu$  ($\equiv  \nu_{k+1}-\nu_k$), of radial modes and $p$ modes 
in terms of the frequency (in mHz) for 
the same model. Finally, the lower panels depict  
the kinetic energy distributions. For periods longer than 
about $1000$ s, the period-spacing distribution of $g$ modes shows 
a regular pattern of mode trapping with a very short trapping 
cycle (period interval between two
trapped modes) of $\sim 100$ s and amplitude, superimposed on  
a high-amplitude variation
characterised by a long trapping cycle 
($\sim 1800$ s). The simultaneous presence of these two
patterns is the result of the double-layered chemical structure 
of the H profile that characterises the  ELM WD template model 
at that effective temperature. This double-layered structure 
becomes a single-layered one by the action of element diffusion
(see Sect. \ref{diffusion}).
The period spacing does not reach 
the asymptotic value in the period interval 
shown in the figure. The $E_{\rm kin}$ distribution only shows the 
long-trapping cycle pattern. We examined the radial 
eigenfunctions
of two modes, one of them with $k= 29$ and associated to a local minimum
of $E_{\rm kin}$, and the other one with $k= 33$, corresponding to a 
maximum $E_{\rm kin}$ value. We realised that the eigenfunction of the 
$k= 33$ mode has larger amplitudes at the core regions 
than the $k= 29$ mode.
It is apparent that as a general rule, all the $g$ modes
in the ELM WD template model are confined to the core regions, but 
some of them have eigenfunctions with larger amplitudes than 
the remaining ones. They correspond  to local maxima in the 
$E_{\rm kin}$ distribution. On the other hand, the frequency 
spacing of radial modes
and $p$ modes are close to the asymptotic value 
for radial order $k \leq 9$ and $k \leq 16$, respectively. In this case, 
no mode-trapping signatures are visible in the kinetic energy 
distribution of the modes.

\begin{figure} 
\begin{center}
\includegraphics[clip,width=9 cm]{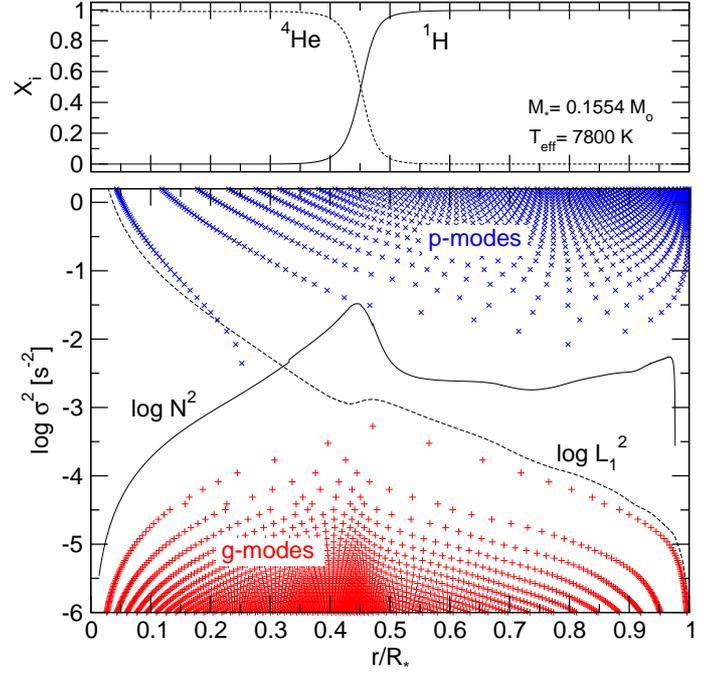} 
\caption{Chemical profiles of  He and H (upper panel) and
the propagation diagram (lower panel) 
for  the ELM WD template  model  of 
$M_*=  0.1554  M_{\odot}$  
and  $T_{\rm eff}  \approx
  7\,800$ K.}
\label{figure-xfbvl-7800-0.1554} 
\end{center}
\end{figure}

The  results for the  LM  WD template  model with  $M_*=
0.2389  M_{\odot}$  and  $T_{\rm  eff}\sim  9\,600$ K  are  depicted  in
Fig. \ref{figure-delp-ekin-10000-0.2389}. For $g$ modes,  
mode-trapping signatures  are quite evident, but the trapping 
cycle ($\sim 200$ s) of the present pattern 
is much shorter than for the high-amplitude variation of the
 ELM WD template model. Mode-trapping features are absent 
from the $E_{\rm kin}$ distribution, which instead is very 
smooth. Note that 
the $\Delta \Pi$ values approach to the asymptotic prediction 
for periods longer than $\sim 4000$ s. The frequency separations of 
radial modes and $p$ modes, 
on the other hand, show some signatures of mode trapping for low-order modes, and do not reach the asymptotic value for the range of 
frequencies considered in the plot.

\begin{figure} 
\begin{center}
\includegraphics[clip,width=9 cm]{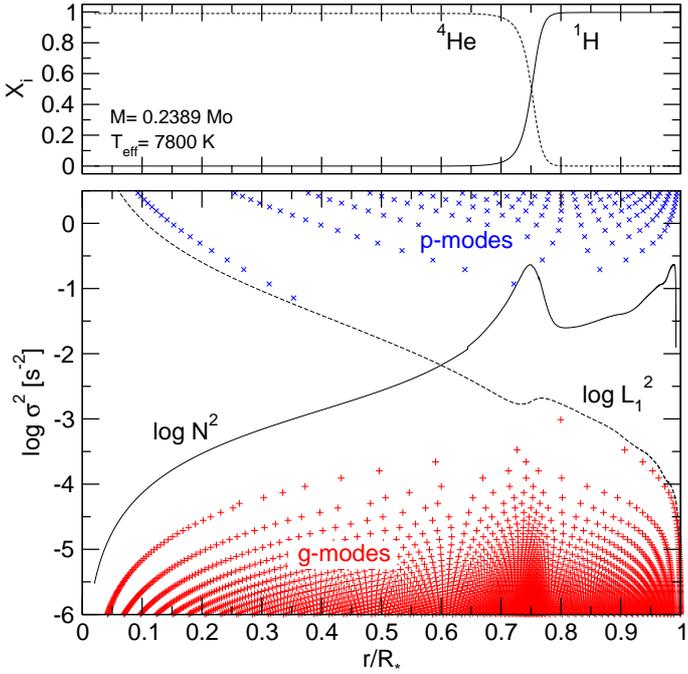} 
\caption{Same as Fig. \ref{figure-xfbvl-7800-0.1554}, but 
for the LM WD template model with $M_*= 0.2389 M_{\odot}$.}
\label{figure-xfbvl-7800-0.2389} 
\end{center}
\end{figure}

We now describe the results for template models located at an
effective temperature close to the $T_{\rm eff}$ of the coolest 
known pulsating low-mass WD (SDSS J222859.93+362359.6,
$T_{\rm eff}= 7870 \pm 120$ K), which defines the empirical red edge 
of the instability strip of this new kind of variable stars. 
Specifically, we analyse two template models 
at $T_{\rm eff} \sim 7800$ K, one of them with a stellar mass  
$M_*= 0.1554 M_{\odot}$, and the other 
one with $M_*= 0.2389 M_{\odot}$. The location of these template
models in the $\log T_{\rm eff} - \log g$ plane is shown 
in Fig. \ref{figure-HR1} with
black squares. In Fig. \ref{figure-xfbvl-7800-0.1554}
and \ref{figure-xfbvl-7800-0.2389} we show their chemical profiles 
(upper panels) and propagation diagrams (lower panels). 
For the ELM template model, 
the changes in the chemical profiles caused by element diffusion 
while the star cools from $T_{\rm eff} \sim 9600$ K to $T_{\rm eff} \sim 7800$ 
are very noticeable. The shape of the He/H interface 
has changed from a 
double-layered to a single-layered structure
(see Sect. \ref{diffusion}). At this lower $T_{\rm eff}$ the He/H transition  is located
at  $r/R_* \sim 0.45$ and results in a more external bump of the 
Brunt-V\"ais\"al\"a frequency than in the model at 
$T_{\rm eff} \sim 9600$ K (compare with Fig. 
\ref{figure-xfbvl-10000-0.1554}). For the LM WD template model, 
the changes in the He/H chemical transition region and in
the Brunt-V\"ais\"al\"a frequency 
due to element diffusion are less noticeable. When we compare
the upper panels of Figs. \ref{figure-xfbvl-7800-0.2389} and \ref{figure-xfbvl-10000-0.2389}, we barely note a slight variation 
in the thickness of the He/H transition region, which is 
somewhat narrow for the cooler model. This results in a slightly 
more narrow bump in the profile of $N^2$.

\begin{figure} 
\begin{center}
\includegraphics[clip,width=9 cm]{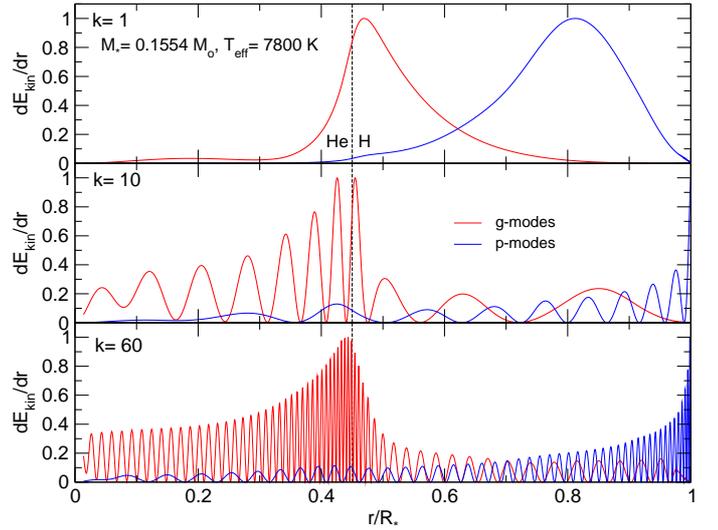} 
\caption{Run  of the density of  kinetic energy
 $dE_{\rm ekin}/dr$ (normalised to 1) for dipole 
$g$  (red) and $p$ modes (solid blue curves) 
with $k= 1$ (upper panel), $k= 10$ (middle panel),  and 
$k= 60$ (lower panel),  for the
ELM template model with $M_*=  0.1554 M_{\odot}$ and
$T_{\rm eff}  \approx 7\,800$ K 
(analysed in Fig.\ref{figure-xfbvl-7800-0.1554}).}
\label{figure-dens-ekin-7800-0.1554} 
\end{center}
\end{figure}

\begin{figure} 
\begin{center}
\includegraphics[clip,width=9 cm]{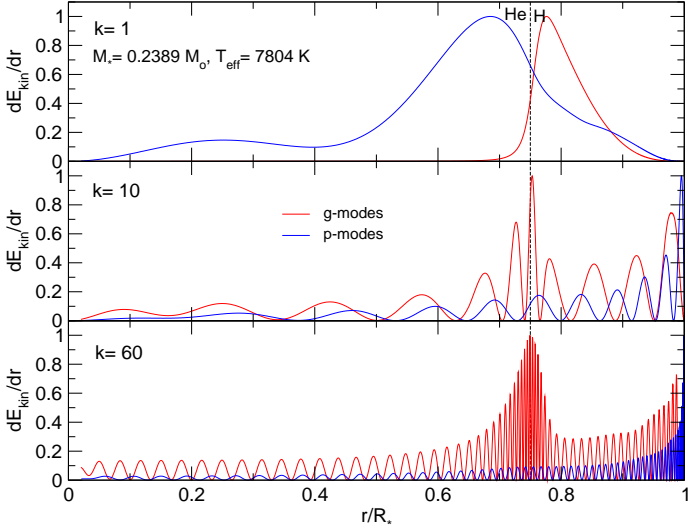} 
\caption{Same as Fig. \ref{figure-dens-ekin-7800-0.1554}, but 
for the LM WD template model with $M_*= 0.2389 M_{\odot}$
analysed in Fig. \ref{figure-xfbvl-7800-0.2389}.}
\label{figure-dens-ekin-7800-0.2389} 
\end{center}
\end{figure}

In Figs. \ref{figure-dens-ekin-7800-0.1554} and 
\ref{figure-dens-ekin-7800-0.2389} we show $dE_{\rm kin}/dr$
for $p$ and $g$ modes\footnote{We do not show the results for radial modes 
because they look very similar to those of $p$ modes.} with 
$k= 1, 10,$ and $60$ for the ELM and the LM template models at $T_{\rm eff} \sim 7800$ K. For the ELM
model there is a clear distinction between $p$ and $g$ modes
for the part of the star that they probe: $g$ modes
are mostly confined to the core,  $p$ modes to the envelope.
The situation is therefore the same as for the ELM WD template model 
at $T_{\rm eff} \sim 9600$ K (see Fig. \ref{figure-dens-ekin-10000-0.1554}). 
A similar situation is found for the LM template model. At $T_{\rm eff} \sim 7800$ the $g$ modes are very sensitive to the 
presence of the He/H transition region, whereas $p$ modes 
(and radial modes, not shown) are concentrated towards the stellar 
surface and are unaffected by the presence of that chemical interface.

Finally, in Figs. \ref{figure-delp-ekin-7800-0.1554}  and
\ref{figure-delp-ekin-7800-0.2389} we show the distributions 
of $\Delta \Pi$ ($g$ modes), $\Delta \nu$ ($p$ modes), 
and $\log(E_{\rm kin})$ for the 
ELM and the LM WD template models at $T_{\rm eff}  \approx 7\,800$ K.
It is evident that the period-spacing pattern of $g$ modes 
of the ELM template model at this low effective temperature is 
completely different from that for the ELM  model 
at high $T_{\rm eff}$. Although the asymptotic period-spacing (and 
consequently, the average period-spacing) has 
not changed much, the $\Delta \Pi$ distribution exhibits only a single
signal of mode trapping, characterised by a short trapping cycle. 
Moreover, at variance with the model at  $T_{\rm eff}  \approx 9\,600$ K,
$\Delta \Pi$ reaches the asymptotic value for $\Pi \gtrsim 4000$ s.
The very distinct pattern of $\Delta \Pi$ in comparison 
with the hot ELM template model is due to the strong 
differences in the chemical profiles, and ultimately in the run of 
the Brunt-V\"ais\"al\"a frequency (compare 
Figs. \ref{figure-xfbvl-10000-0.1554} and 
\ref{figure-xfbvl-7800-0.1554}). The frequency-spacing pattern for  
$T_{\rm eff}\sim 7800$ K of the p modes (and radial modes, not
shown) is similar to that found
for  $T_{\rm eff}\sim 9600$ K, but there is a strong difference in the 
asymptotic frequency-spacing (and thus, in the average frequency-spacing).
In fact, $\Delta \nu^{\rm a}= 4.16$ mHz for the cool model is
about three times larger than for the hot model 
($\Delta \nu^{\rm a}= 1.51$ mHz). This marked difference
 originates in the fact that for the model at 
$T_{\rm eff}\sim 7800$ K the Lamb frequency adopts much higher values
than for the model at $T_{\rm eff}\sim 9600$ K, and in turn, 
the integral in Eq. (\ref{afs}) is lower, and consequently, 
$\Delta \nu$ is higher. 

We did not find  
substantial differences in the patterns of $\Delta \Pi$ and 
$\Delta \nu$ (Fig. \ref{figure-delp-ekin-7800-0.2389}) in the
LM\ WD template model  compared with 
the hot template model 
(Fig. \ref{figure-delp-ekin-10000-0.2389}). For the 
cool template model, both $\Delta \Pi^{\rm a}$ and 
$\Delta \nu^{\rm a}$, and in turn the average $\Delta \Pi$ and 
$\Delta \nu$, are larger than for the hot template model, however.
This is a direct consequence of the decrease in the 
Brunt-V\"ais\"al\"a frequency and the increase in the Lamb frequency  
for decreasing effective temperatures. 
This behaviour was anticipated in Sect. \ref{figure-asintoticos} 
where we discussed the dependence of $\Delta \Pi^{\rm a}$ and 
$\Delta \nu^{\rm a}$ on the effective temperature.

\subsection{Effects of the total mass and the effective temperature}
\label{effect-mass-teff}

The period spacing  and the
periods themselves  of $g$ modes vary as  the  inverse of  
the Brunt-V\"ais\"al\"a  
frequency (see Sect.  \ref{asymptotic}).
The Brunt-V\"ais\"al\"a  frequency, in turn, 
increases with  higher stellar masses and  with higher effective  
temperatures.  As  a result,  the pulsation periods of $g$ modes 
are  longer  for  smaller  mass  (lower  gravity)  
and  lower effective  temperature  (increasing degeneracy).   For 
 $p$-modes and radial modes the sound speed decreases for lower $M_*$ and 
higher $T_{\rm eff}$, and consequently, 
the periods increase. We study the effects of $M_*$ and 
$T_{\rm eff}$ separately below.

The  effects of  the
stellar   mass   on  the   pulsation   periods   is   shown  in   Fig.
\ref{figure-per-mass-l0l1}, where we plot $\Pi$  for non-radial $\ell= 1$ 
$g$ modes and $p$ modes, and also for radial 
modes ($\ell= 0$) in terms of $M_*$ for 
a fixed value of $T_{\rm  eff}= 9\,500$ K. 
The periods of the three types of eigenmodes increase with decreasing 
stellar mass, as expected. The
period gap that separates the families of $p$ modes/radial modes 
and  $g$ modes notoriously shrinks for ELM WD models.
The  striking step in the run of the periods in terms of the mass 
is associated  with the limit stellar mass value, at $\sim 0.18 M_{\odot}$.
Note the strong increase of the $p-$  and 
radial-mode periods for ELM WDs ($M_* \lesssim 0.18 M_{\odot}$).
As can be seen, the period of the 
fundamental radial mode ($r_0$) is substantially longer
than the period corresponding to the $k= 1$ $p$ mode ($p_1$) for LM WDs 
($M_* \gtrsim 0.18 M_{\odot}$), but both periods 
adopt virtually the same values (as do other low-order pairs of 
modes, as well: $r_1$-$p_2$, $r_2$-$p_3$, etc) for ELM WD models. 
Figure \ref{figure-per-mass-l2}
displays the results for  $g$ , $f$ and $p$ modes with 
$\ell= 2$. In this case, the $g-$ mode periods are substantially
shorter than $\ell= 1$, and  the 
branches of $g$ and $p$ modes are clearly split by the period 
of the $f$ mode.

\begin{figure} 
\begin{center}
\includegraphics[clip,width=9 cm]{delp-ekin-7800-0.1554.eps} 
\caption{ $\ell= 1$ forward period-spacing  of $g$ modes versus 
periods (upper left-hand panel), and the forward frequency-spacing of 
radial modes (hollow small circles) and $p$ modes (filled 
small circles) versus frequencies (upper right-hand panel) and the 
corresponding kinetic energy distributions (lower panels) for 
the ELM WD template model with $M_*= 0.1554  M_{\odot}$ and 
$T_{\rm eff}\sim  7\,800$ K.}
\label{figure-delp-ekin-7800-0.1554} 
\end{center}
\end{figure}

\begin{figure} 
\begin{center}
\includegraphics[clip,width=9 cm]{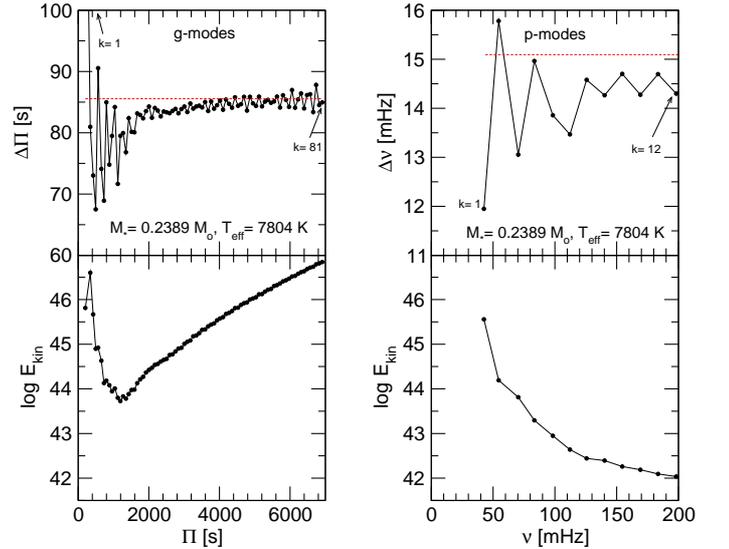} 
\caption{Same as Fig.\ref{figure-delp-ekin-7800-0.1554}, but for 
the LM WD template model with $M_*= 0.2389 M_{\odot}$.}
\label{figure-delp-ekin-7800-0.2389} 
\end{center}
\end{figure}

In summary, the deep structural differences between LM and ELM WD models  
are clearly illustrated by the very different behaviour of the 
periods (in particular of $p$ modes and radial modes), as documented 
in Figs.  \ref{figure-per-mass-l0l1} and \ref{figure-per-mass-l2}.

In Fig.  \ref{figure-per-teff}
we show the evolution of  the pulsation periods  
of $\ell= 1$ $g$ and $p$ modes, and also radial 
modes with $T_{\rm eff}$ for models with $M_*= 0.1917 M_{\odot}$.  The lengthening of $g-$ mode periods with decreasing effective temperature is evident,  
although the effect is much weaker than  the decrease with the 
stellar mass (compare with Fig. \ref{figure-per-mass-l0l1}).
The $p-$  and radial-mode periods, on the other hand, decrease with 
decreasing $T_{\rm eff}$. 

\begin{figure} 
\begin{center}
\includegraphics[clip,width=9 cm]{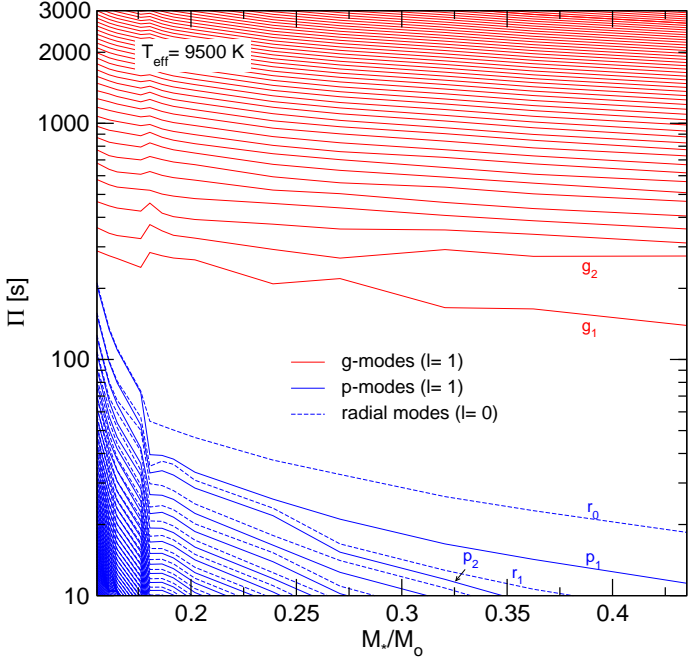} 
\caption{Pulsation  periods of  $\ell= 1$  $g$- and $p$ modes and also 
radial modes ($\ell= 0$) in  terms  of the
  stellar  mass  for $T_{\rm  eff}=  9500$  K.  Periods increase  with
  decreasing $M_*$.}
\label{figure-per-mass-l0l1} 
\end{center}
\end{figure}

\begin{figure} 
\begin{center}
\includegraphics[clip,width=9 cm]{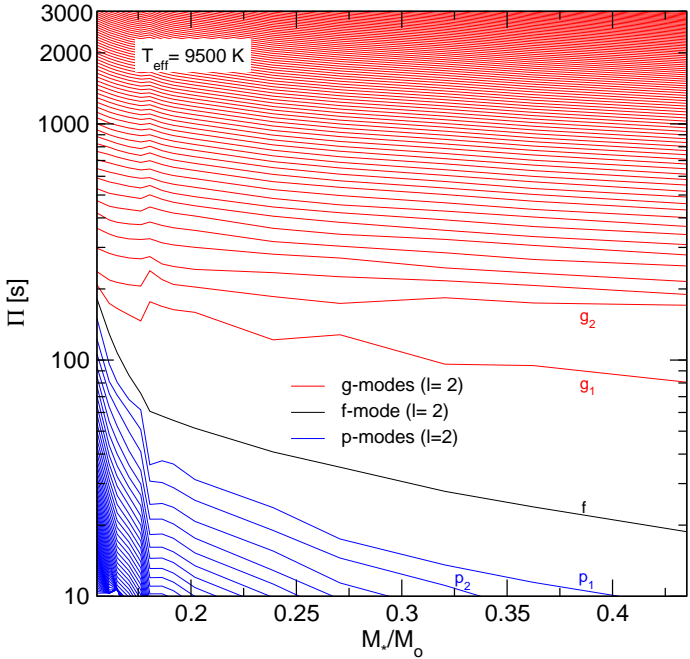} 
\caption{Same as Fig. \ref{figure-per-mass-l0l1}, but 
for pulsation  periods of  $\ell= 2$  $g$, $f,$ and $p$ modes.}
\label{figure-per-mass-l2} 
\end{center}
\end{figure}

\subsection{Effects of element diffusion}
\label{diffusion}

Here, we  describe the effects of the evolving  chemical profiles 
on the  pulsation properties of  low-mass, He-core WDs.  This matter
has been extensively explored in previous papers 
\citep[see][in the context of   DAV stars]{2002MNRAS.332..392C}.
Time-dependent element diffusion strongly  modifies the
shape of  the He  and H  chemical profiles as  the WD cools,
causing H to float to the surface and He to sink down.  In particular,
diffusion  not only  modifies the  chemical composition  of  the outer
layers,  but also  the shape  of the  He/H chemical  transition region
itself.  This is clearly  documented  in Fig.  \ref{figure-difu-0.1554}
for the $0.1554 M_{\odot}$ ELM  sequence in the  $T_{\rm eff}$
interval ($9600-8000$ K).  For  the model at $T_{\rm eff}= 9600$
K,   the   H  profile   is   characterised   by  a   diffusion-shaped
double-layered chemical structure, which consists of a pure H envelope
 on top of an  intermediate remnant  shell rich in  H and He.
This structure still  remains, although to a much  weaker extent, in the
model at  $T_{\rm eff}= 9000$  K.  Finally,  from $T_{\rm
  eff}\sim  8040$ K down, the H profile adopts a single-layered 
chemical structure.  This type of transitions of the shape of chemical 
profiles caused by element diffusion has previously been studied in detail
in DB WDs \citep[see, e.g.,][]{2004A&A...417.1115A}. Element diffusion processes affect  all the sequences
considered   in   this  paper,   although   the   transition  from   a
double-layered  to a single-layered structure  occurs at different
effective  temperatures.  The  low  surface  gravity  that
characterises  the model shown in Fig.  \ref{figure-difu-0.1554} 
($\log g \sim 5.5-6.2$), which  results in  a weak  impact of
gravitational settling, and  the very long timescale that characterises
element diffusion processes at the depth where the He/H interface 
is located ($r/R_* \sim 0.2$), eventually  lead to a wider chemical 
transition than for more massive models.  Because of  this, the
sequences  with larger masses ($M_* \gtrsim 0.19 M_{\odot}$) 
reach  the single-layered configuration at
effective temperatures higher than $\sim 10\,000$ K,  
beyond the domain in which pulsating objects are currently found.

\begin{figure} 
\begin{center}
\includegraphics[clip,width=9 cm]{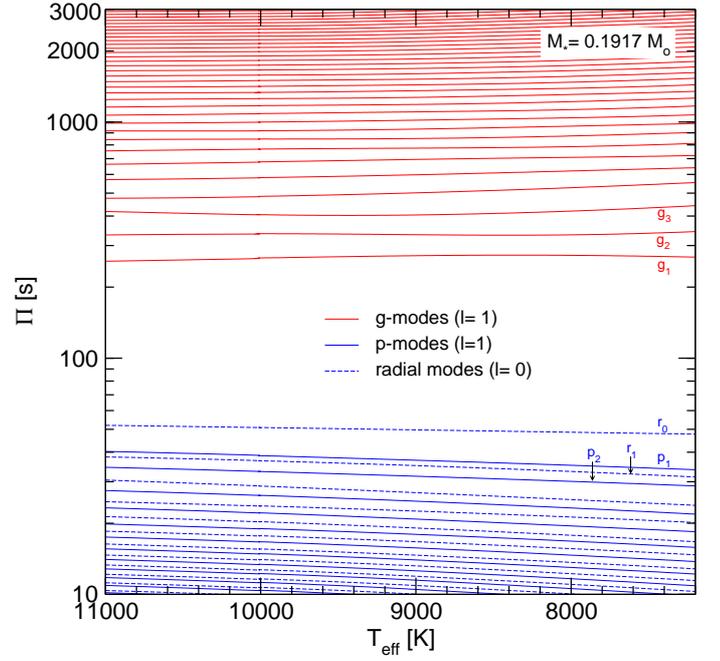} 
\caption{Pulsation  periods of  $\ell= 1$  $g$ modes and 
$p$ modes/radial modes in  terms  of the effective  temperature for 
$M_*=  0.1917 M_{\odot}$.   $g-$ mode periods increase
  with decreasing $T_{\rm eff}$; the opposite holds for 
$p$ modes and radial modes.}
\label{figure-per-teff} 
\end{center}
\end{figure}

The evolution of the  shape of the  He/H chemical transition region 
are  translated into noticeable  changes  in  the  run of  the  
Brunt-V\"ais\"al\"a frequency (lower panel of Fig. \ref{figure-difu-0.1554}),  
as a result of the changing Ledoux term $B$ (middle 
panel of Fig. \ref{figure-difu-0.1554}).  In  fact, at  high
effective  temperatures, $N^2$  is characterised  by two  bumps 
(the most prominent one located at $r/R_* \sim 0.25$ and the smaller at $r/R_* \sim 0.62$), which
merge  into  a  single bump  (at $r/R_* \sim 0.5$) when  the  
chemical  interface  He/H adopts  a  single-layered  structure.  
As was first emphasised by \citet{2012A&A...547A..96C}, 
and in  the light of
our results, diffusive equilibrium is not a valid assumption in 
the He/H  transition region for ELM WD stars. 
Exploratory computations without element diffusion 
result in changes of up to 5 \% in the value of the periods 
compared with computations with element diffusion.

The effects of element diffusion for a more massive model
(an LM WD model with $M_*= 0.2389 M_{\odot}$) 
at the same range of effective temperatures is less impressive,
as is shown in Fig. \ref{figure-difu-0.2389}. In this case, the 
H chemical profile does not change dramatically, and the same 
occurs with $B$ and the Brunt-V\"ais\"al\"a frequency. Even so, this
slight evolution in the shape of the chemical profiles  
is translated into non-negligible changes in the pulsation periods. Again, 
for LM WD stars it is necessary to properly account for element 
diffusion processes to accurately compute the period spectrum.

\subsection{Template models with $M_*$ near the threshold mass}
\label{threshold}

It is interesting to examine the pulsation properties of 
low-mass WD sequences with stellar masses near the critical
mass for the development of CNO flashes ($M_∗ \sim 0.18 M_{\odot}$). 
This is because at least three of the five ELM
pulsating WDs reported by \citet{2013MNRAS.436.3573H} have
stellar masses near this threshold value (see Fig. \ref{figure-HR1}). 
Therefore it might be quite important to find an asteroseismic prospect 
to distinguish ELM from LM WDs \citep{2013A&A...557A..19A}.
Specifically,  we focus 
on the most massive ELM WD sequence ($M_* = 0.1762 M_{\odot}$), and
the lowest mass LM WD sequence ($M_* = 0.1805 M_{\odot}$). In 
Figs. \ref{figure-xfbvldekin-10000-0.1762} and 
\ref{figure-xfbvldekin-10000-0.1805} we display the chemical 
profiles (upper panels) and the propagation diagrams (central panels)
for two models at $T_{\rm eff} \sim 10\,000$ K. 
Interestingly enough, the H content and the shape of 
the chemical  interface of He/H are very 
different even though the difference in the stellar 
mass of these two models is almost negligible: 
$\Delta M_*= 0.0043 M_{\odot}$. The differences in the shape and
location of 
the He/H interface (by virtue of the different thicknesses of the 
H envelope) for both models are translated into distinct features 
in the run of the squared critical frequencies, in particular in the 
Brunt-V\"ais\"al\"a frequency (middle panels). Note, 
in particular, the presence of two bumps in $N^2$ 
in the  LM model with $M_* = 0.1805 M_{\odot}$
, which result from a double-layered chemical structure at 
the He/H interface. It differs from the single bump of 
$N^2$ for the ELM model with $M_* = 0.1762 M_{\odot}$, 
which results from the single-layered shape 
of the He/H chemical transition region.

\begin{figure} 
\begin{center}
\includegraphics[clip,width=9 cm]{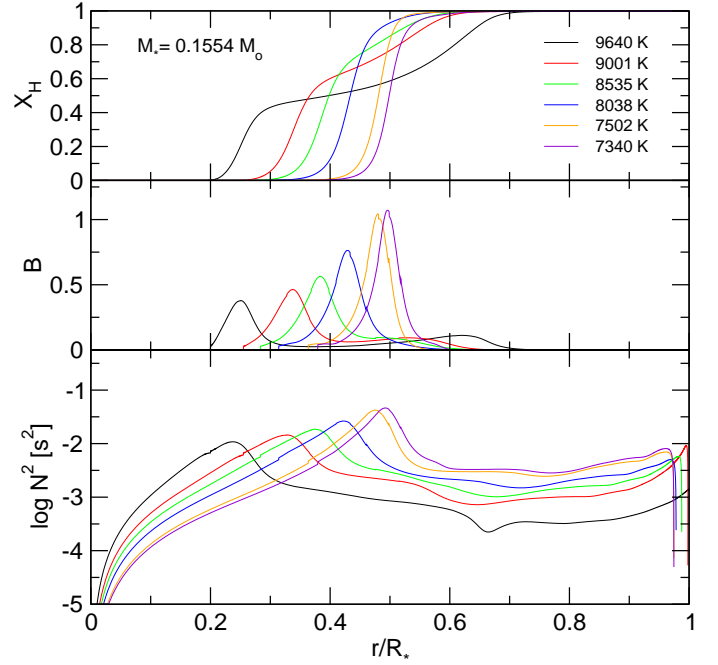} 
\caption{Internal chemical profile of  H (upper panel), 
the Ledoux term $B$
entering in  the computation of the Brunt-V\"ais\"al\"a frequency
[see Eq.(\ref{BLedoux})] (middle panel), and the logarithm
  of the squared Brunt-V\"ais\"al\"a frequency (lower panel), 
for an ELM WD model with $M_*=
  0.1554  M_{\odot}$  at  different  effective temperatures, as indicated.}
\label{figure-difu-0.1554} 
\end{center}
\end{figure}

The consequences  of these differences are clearly illustrated in the
propagation  characteristics of the pulsation modes. The lower panels of
Figs.  \ref{figure-xfbvldekin-10000-0.1762} and
\ref{figure-xfbvldekin-10000-0.1805} display the kinetic
energy density of pulsation for $\ell= 1$ $g$ and $p$ modes with $k= 1, 10,$
and $20$.  As we found in Sect. \ref{template}, for the ELM WD model
$g$ modes are confined to the He core, and $p$ modes have most of
their kinetic energy located at the surface regions
(Fig. \ref{figure-xfbvldekin-10000-0.1762}). For $g$ modes in the LM\ WD model, the kinetic energy is distributed throughout
the model, and they are sensitive to the presence of the He/H
interface.  $p$ modes, on the other hand, are rather insensitive to
the chemical gradient  and have most of their kinetic energy placed at
the stellar surface (Fig.  \ref{figure-xfbvldekin-10000-0.1805}). 

In Fig. \ref{figure-delp-delf-10000-0.1762-0.1805} we show the forward
period-spacing in terms of the dipole $g-$ mode  periods (upper panels)
and the forward frequency-spacing in terms  of the dipole $p$
mode
frequencies (lower panels) for the  ELM WD template model with $M_*=
0.1762 M_{\odot}$  (left) and the LM WD template model with $M_*=
0.1805 M_{\odot}$ (right), the same models as considered  in Figs.
\ref{figure-xfbvldekin-10000-0.1762} and
\ref{figure-xfbvldekin-10000-0.1805}.  Mode-trapping features in the form of strong departures of
uniform period-spacing of $g$ modes  can be appreciated in both
models, in particular for  periods shorter than about $2000$ s for the  ELM WD model, and $\Pi \lesssim 4000$ s for the LM WD
model.  Longer periods approach the asymptotic period-spacings,
although even with low-amplitude deviations because of mode trapping.  For periods shorter than $\sim 3000$ s, the
amplitudes of deviations from a constant period  separation are
markedly larger for the LM than for the  ELM model. In principle, this difference might be considered as a practical tool for distinguishing
stars with CNO flashes in their early-cooling phase
from those without, provided that enough consecutive low- and
intermediate-order $g$ modes with the same harmonic degree       were
observed in pulsating low-mass WDs \citep{2013A&A...557A..19A}. 
This optimistic view is somewhat
dampened because the most useful low radial-order $g$ modes
are most likely pulsationally stable, as shown from   the non-adiabatic
analysis of \citet{2012A&A...547A..96C} and demonstrated  by most of 
the observed pulsating low-mass WD stars, in which  no periods 
shorter than about 1400 s are seen.

\begin{figure} 
\begin{center}
\includegraphics[clip,width=9 cm]{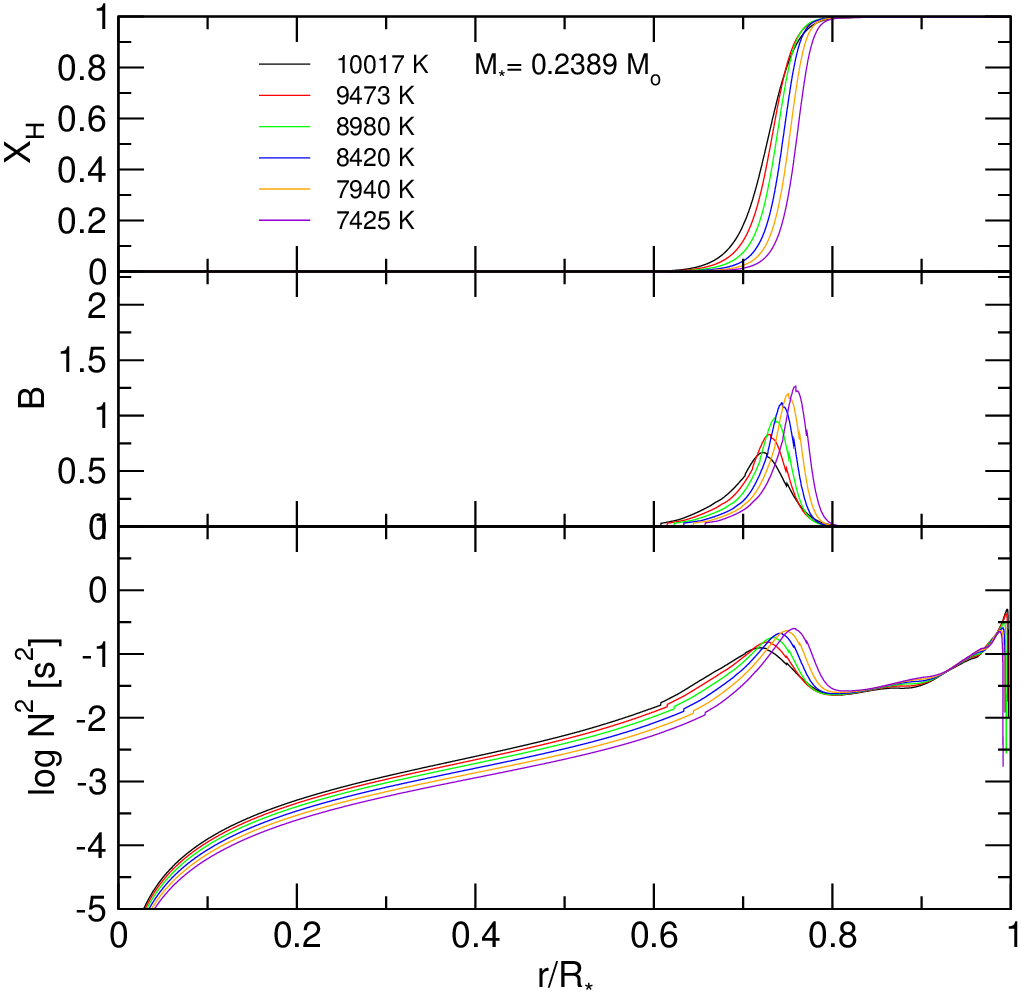} 
\caption{Same as Fig. \ref{figure-difu-0.1554}, but for an LM WD model
with $M_*= 0.2389  M_{\odot}$.}
\label{figure-difu-0.2389} 
\end{center}
\end{figure}

Finally, as for the $p$-modes, there are  clear
non-uniformities in the frequency spacing distributions,  which are of
similar amplitudes in both WD models.  However, the asymptotic
frequency-spacing for the LM WD model  is about twice that for the ELM
WD model, even though  the difference in stellar mass is
only $\sim 0.004 M_{\odot}$.  This strong difference in the mean
frequency-spacing  could also be exploited to distinguish between
the two types of  objects. To be able to employ this property as a
useful seismological tool, however, it will be necessary to detect many
$p$ mode consecutive periods in pulsating low-mass WDs. At present,
only a few short periods ($\Pi \sim 108-134$ s)  have been detected in
only one low-mass pulsating WD (SDSS J111215.82+111745.0), and still
it remains to be determined  whether they are genuine $p$ modes or not
(see Sect. \ref{p-modes-real-star} ). Thus, this potential
asteroseismological tool is for now only of academic interest.

\section{Interpretation of observations}
\label{observations}

Until today, five pulsating low-mass WD stars have been detected: SDSS  J184037.78+642312.3 
\citep{2012ApJ...750L..28H}, SDSS  J111215.82+111745.0, SDSS
J151826.68+065813.2 \citep{2013ApJ...765..102H}, 
SDSS  J161431.28+191219.4,  and  SDSS J222859.93+362359.6 
\citep{2013MNRAS.436.3573H}. 
They have very long pulsation periods, 
from $\sim 1180$ s up to $\sim 6240$ s,  although one object (SDSS
J111215.82+111745.0) also
has short-period pulsations ($108-134$ s). Table 4 
of \citet{2013MNRAS.436.3573H} lists the main properties of the five known 
pulsating low-mass WDs. These authors discuss the common 
properties of this set of pulsators at some length. In Fig. \ref {figure-HR1} we display the 
location of the stars in the $\log T_{\rm eff} - \log g$ diagram, while  
in Figs. \ref{figure-periods-teff-observed} and 
\ref{figure-periods-gravity-observed} we show the period 
spectrum of these pulsating stars in terms of the effective temperature
and surface gravity. The long-period pulsations are 
most likely the result of intermediate- and high-order $g$ modes 
($12 \lesssim k \lesssim 50$ for $\ell= 1$) excited by the 
$\kappa-\gamma$-mechanism that acts in the H partial ionisation 
zone, as indicated by stability computations 
\citep{2012A&A...547A..96C,2013ApJ...762...57V}. On the other 
hand, the short periods detected in SDSS
J111215.82+111745.0 might be caused by $p$ modes or even radial modes 
of low radial order, as was suggested by stability computations,
although the precise nature of these periodicities 
still remains to be defined\footnote{We recall that
the vertical displacements of $p$ modes and radial modes
are much smaller than the horizontal displacements 
of $g$ modes caused by the high gravity, which hinders detecting
this type of modes even in low-mass WDs ($\log g\sim 6$).}. 

\begin{figure} 
\begin{center}
\includegraphics[clip,width=9 cm]{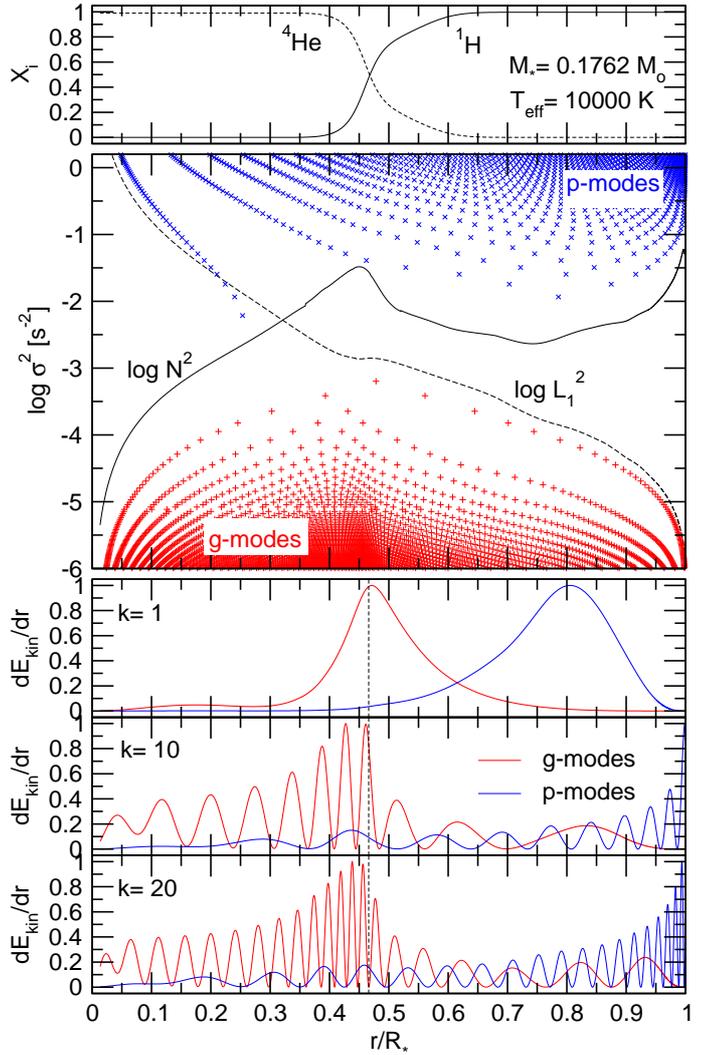} 
\caption{Chemical profiles of  He and H (upper panel),
the propagation diagram (centre panel), and the 
kinetic energy density for dipole 
$g$ (red) and $p$ modes (solid blue curves) 
with $k= 1, 10,$ and 20 (lower panels),  
for   an ELM WD model  with 
$M_*=  0.1762  M_{\odot}$  and  $T_{\rm eff}  \approx
  10\,000$ K.}
\label{figure-xfbvldekin-10000-0.1762} 
\end{center}
\end{figure}

\begin{figure} 
\begin{center}
\includegraphics[clip,width=9 cm]{xfbvldekin-10000-0.1805.eps} 
\caption{Same as Fig. \ref{figure-xfbvldekin-10000-0.1762}, but for an 
LM WD model with $M_*=  0.1806  M_{\odot}$.}
\label{figure-xfbvldekin-10000-0.1805} 
\end{center}
\end{figure}

\begin{figure} 
\begin{center}
\includegraphics[clip,width=9 cm]{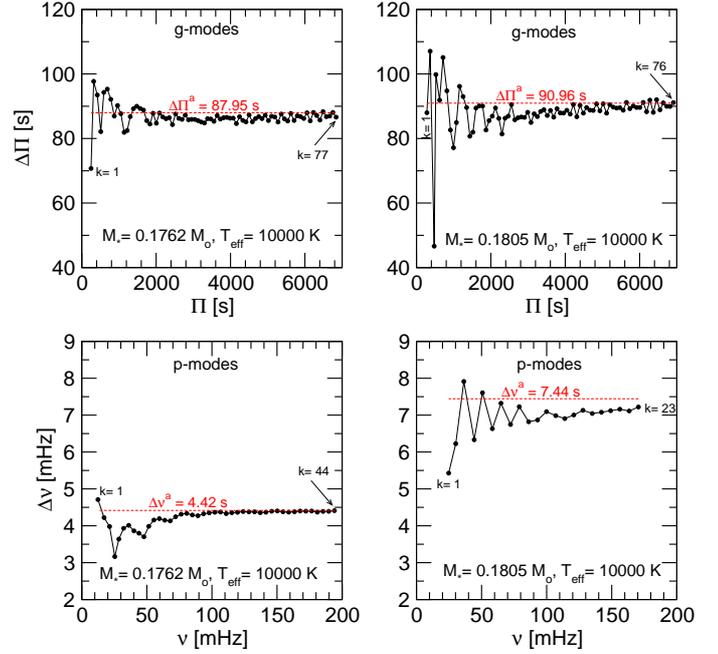} 
\caption{Upper panels show the $\ell= 1$ forward period-spacing of 
$g$ modes vs periods, and lower panels depict the 
forward frequency-spacing of $p$ modes vs frequency. Left-hand 
panels correspond to an ELM WD model with 
$M_*= 0.1762 M_{\odot,}$ right-hand panels to 
an LM WD model with $M_*= 0.1805 M_{\odot}$, both at 
$T_{\rm eff} \sim 10\,000$ K. Red dashed lines correspond to 
the asymptotic predictions.}
\label{figure-delp-delf-10000-0.1762-0.1805} 
\end{center}
\end{figure}

Although there 
are very few objects currently available to establish any trend, Figs. 
\ref{figure-periods-teff-observed} and \ref{figure-periods-gravity-observed}
reveal a certain correlation of the long periods (those presumably 
associated with $g$ modes) with $g$ and $T_{\rm eff}$. Specifically,  
periods are longer for lower gravities and effective temperatures.
These trends are expected from theoretical grounds, as discussed 
by \citet{2013MNRAS.436.3573H}. For instance, in the context of 
the $\kappa-\gamma$-mechanism of mode driving, the trend of the periods 
to be longer for cooler stars is related to the fact that the 
outer convection zone in a WD star deepens with cooling, 
with the consequent increase in the thermal timescale ($\tau_{\rm th}$) 
at its base. Since the periods of the excited modes are of the order of
$\tau_{\rm th}$, modes with increasingly longer periods are 
gradually excited as the WD cools. This is an often-studied 
property of ZZ Ceti stars \citep{2006ApJ...640..956M}\footnote{Note that 
the same generic trend of periods with $T_{\rm eff}$ is predicted
in the frame of the convective driving mechanism, proposed 
by \citet{1990MNRAS.246..510B}, since in this case the 
critical timescale is the 
convective response timescale $\tau_{\rm C}$, which itself
is some multiple of $\tau_{\rm th}$ \citep{2008ApJ...678L..51M}.}. 
Similarly, the trend of periods with gravity (and thus, with $M_*$)
can be understood by realising that lower $g$ imply lower mean densities
($\rho$), and that the pulsation periods roughly scale with the 
dynamical timescale for the whole star, $\Pi \propto \rho^{−1/2}$
\citep{2013MNRAS.436.3573H}.

\begin{figure} 
\begin{center}
\includegraphics[clip,width=9 cm]{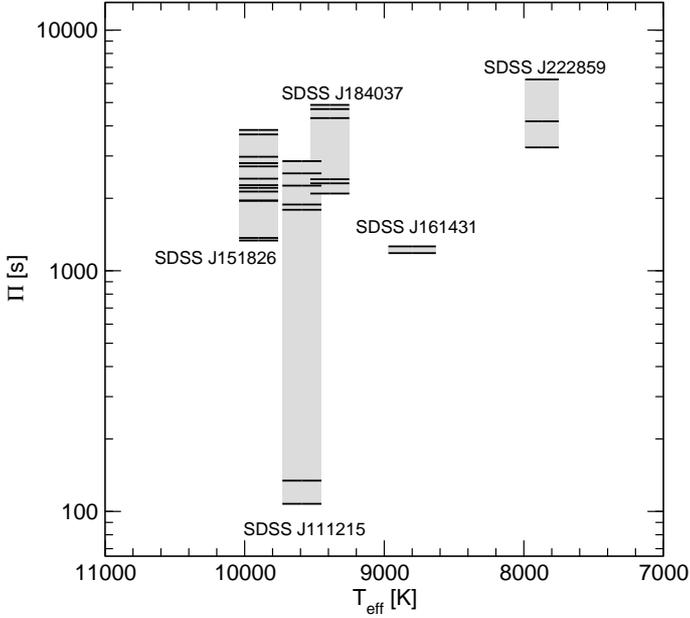} 
\caption{Pulsation periods of the five known pulsating 
low-mass WD stars in terms of the effective temperature, according to 
Table 4 of \citet{2013MNRAS.436.3573H}.}
\label{figure-periods-teff-observed} 
\end{center}
\end{figure}

\begin{figure} 
\begin{center}
\includegraphics[clip,width=9 cm]{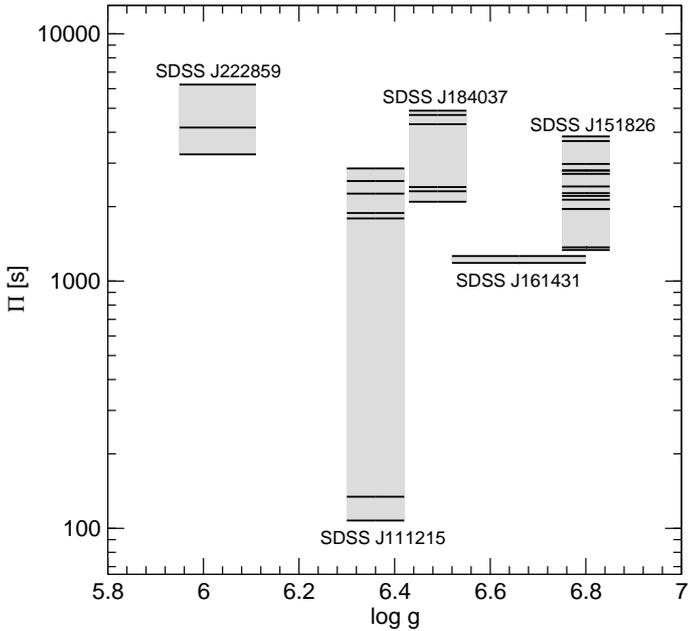} 
\caption{Pulsation periods of the five known pulsating 
low-mass WD stars in terms of the surface gravity, according to 
Table 4 of \citet{2013MNRAS.436.3573H}.}
\label{figure-periods-gravity-observed} 
\end{center}
\end{figure}

We emphasise that while these arguments qualitatively explain the trends
observed in the period spectrum of the five known pulsating low-mass WDs, 
the final decision on this matter relies on detailed non-adiabatic pulsation 
calculations like those performed by \citet{2012A&A...547A..96C} and 
\citet{2013ApJ...762...57V}. We defer a complete and detailed study of 
the non-adiabatic pulsation properties of our set of low-mass He-core WD 
models to a forthcoming paper. In the next two sections, we interpret the short periods in  SDSS
J111215.82+111745.0 by considering adiabatic periods alone 
(Sect. \ref{p-modes-real-star}), and the possibility that
 SDSS J222859.93+362359.6 is a pre-WD star instead of a genuine ELM WD
by using adiabatic computations and some exploratory 
 non-adiabatic results (Sect. \ref{genuine}).

\subsection{ELM WD J111215.82+111745.0: 
the first $p$ modes/radial modes detected in a pulsating WD star?}
\label{p-modes-real-star}

Notwithstanding many theoretical predictions 
\citep{1950ApJ...111..611L,1969ApJ...155..987O,1969Ap&SS...5..113C,1971IAUS...42..145V,1971ApL.....9..161V,1983ApJ...269..645S,1983ApJ...265..982S,1993ApJ...404..294K} and several observational efforts 
\citep{1984AJ.....89.1732R,1994AJ....107..298K,2011A&A...525A..64S,
2013A&A...558A..63C,2014MNRAS.437.1836K}, no $p$ modes or radial 
modes have ever been detected so far in a pulsating WD star of any 
kind. Thus, the discovery of short-period 
pulsations in an ELM WD  by \citet{2013MNRAS.436.3573H} appears to be 
the first detection of these elusive types of pulsation 
modes in a WD star and need to be confirmed. Here, we examine
whether our low-mass He-core WD models are able to 
account for the short-period pulsations observed in J111215.82+111745.0 
and explore the possibility that they might be genuine 
$p$ modes and/or radial modes. 

\begin{figure} 
\begin{center}
\includegraphics[clip,width=9 cm]{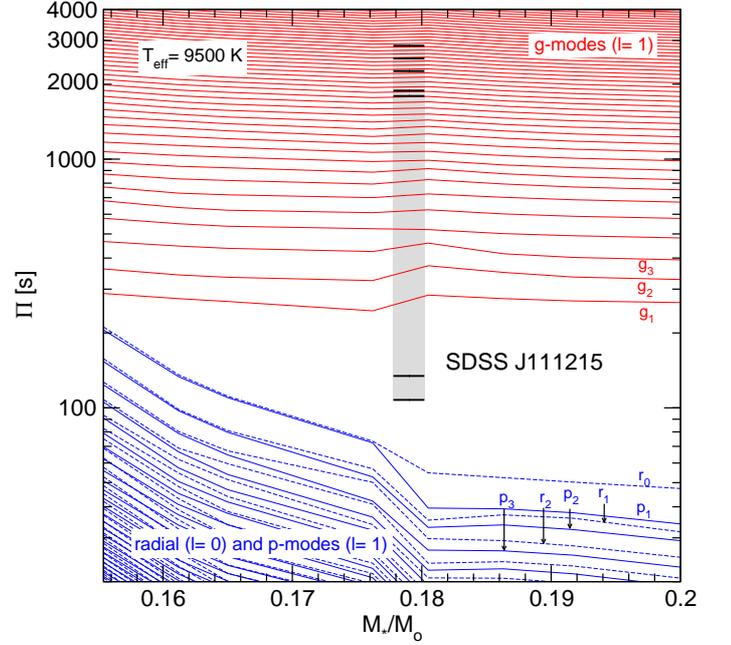} 
\caption{Pulsation periods of our set of low-mass He-core 
WD models in terms of the stellar mass, for radial (blue dashed lines),
non-radial $p$-modes (blue solid lines) and $g$ modes (red solid lines)
at $T_{\rm eff}$, roughly the effective temperature measured for 
J111215.82+111745.0. Horizontal black segments (framed in a grey rectangle) 
represent the periods exhibited by this pulsating star.}
\label{figure-per-mass-sdssj1112-l0l1} 
\end{center}
\end{figure}

\begin{figure} 
\begin{center}
\includegraphics[clip,width=9 cm]{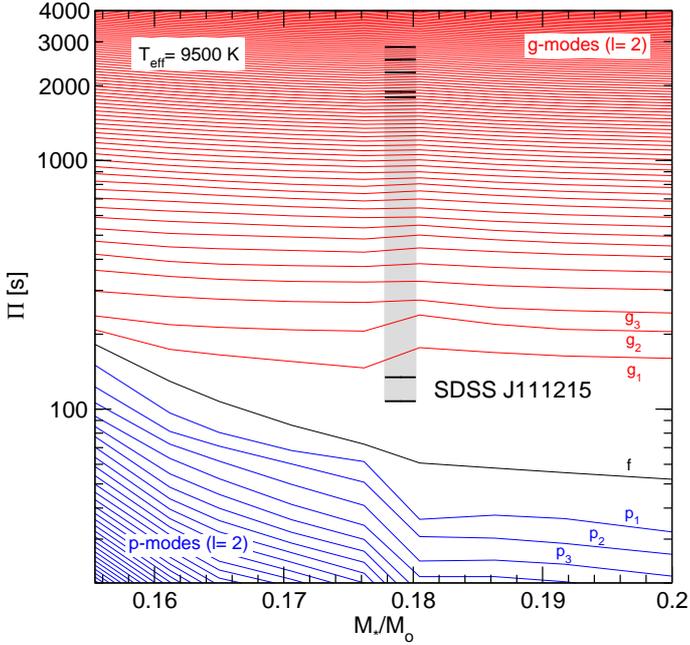} 
\caption{Same as Fig. \ref{figure-per-mass-sdssj1112-l0l1}, but for 
non-radial $\ell= 2$ $p$, $f$, and $g$ modes.}
\label{figure-per-mass-sdssj1112-l2} 
\end{center}
\end{figure}

In Fig. \ref{figure-per-mass-sdssj1112-l0l1} we plot the pulsation
periods of our set of models in terms of $M_*$,  for 
radial modes ($\ell= 0$) and  non-radial $p$ and $g$ modes  with
$\ell= 1$ at roughly the spectroscopic effective temperature derived for
J111215.82+111745.0 ($T_{\rm eff}= 9590 \pm 140$ K). The mass assumed
for this star, $M_*= (0.179 \pm 0.0012) M_{\odot}$  is that  derived
by \citet{2013A&A...557A..19A}  according to the spectroscopic estimation of
the surface gravity  ($\log g= 6.36 \pm 0.06$) as quoted by
\citet{2013MNRAS.436.3573H} (their Table 4).  The periods exhibited 
by this star \citep[107.5600 s, 134.2750 s, 1792.905 s,  1884.599 s, 
2258.528 s, 2539.695 s, 2855.728 s;][]{2013ApJ...765..102H}
are indicated by horizontal black
segments. Clearly, the long-period  pulsations are well accounted for
by high radial-order $g$ modes ($18 \lesssim k \lesssim 30$). However,
at the effective temperature and mass (gravity) of SDSS
J111215.82+111745.0 as predicted by spectroscopy,  our models are 
unable to explain the presence of the short periods.  They lie in the forbidden region in between $g$ and $p$ modes.   An alternative
might be to consider modes with higher harmonic  degree. In
Fig. \ref{figure-per-mass-sdssj1112-l2} we show the results for $\ell= 2$. Even in this  case, the short
periods exhibited by SDSS J111215.82+111745.0 are not accounted for by
our models, although now they are very close to  the lowest-order
$g-$ mode periods. We did not perform pulsation  computations
for  higher harmonic degrees, but we expect that the
periods at 107.5600 s, 134.2750 s can readily been  accounted for by
low-order $g$ modes with $\ell= 3$. In this case,  however, it would
be difficult to conceive the detection   of $\ell= 3$ modes due to
geometric cancellation effects  \citep{1977AcA....27..203D}. If we relax
the constraint of the stellar  mass (gravity), these short periods
might be attributed to low-order $p$ modes and/or radial modes,  if
the stellar mass were somewhat lower ($M_* \sim 0.16 M_{\odot}$).
Alternatively, they might be associated with low-order $g$ modes if the
stellar  mass were substantially larger ($M_* \sim 0.43
M_{\odot}$). Finally,  we might relax the constraint imposed by the
effective temperature  to determine whether we can accommodate the short periods
observed with  theoretical periods. The width of the
gap between  $p$  and $g$ modes decreases slightly for higher
effective temperatures (see Fig. \ref{figure-per-teff}). However,
since the sensitivity of the  periods with the effective temperature
is by far weaker than  with the stellar mass (see
Sect. \ref{effect-mass-teff}),  the change in the periods  (for
reasonable variations of $T_{\rm eff}$) is insufficient for the
theoretical periods of low-order ($k\sim 1-3$)  $g$ and $p$ modes, and
radial  modes be similar to the observed ones. 

In summary, if the temperature and mass (gravity) of 
SDSS J111215.82+111745.0 are correct, our models are unable to explain the short periods, and in particular, they cannot be 
attributed to $p$ modes and/or radial modes.

\subsection{ SDSS J222859.93+362359.6: an ELM WD or  
a pre-WD star?}
\label{genuine}

\begin{figure*} 
\begin{center}
\includegraphics[clip,width=13 cm]{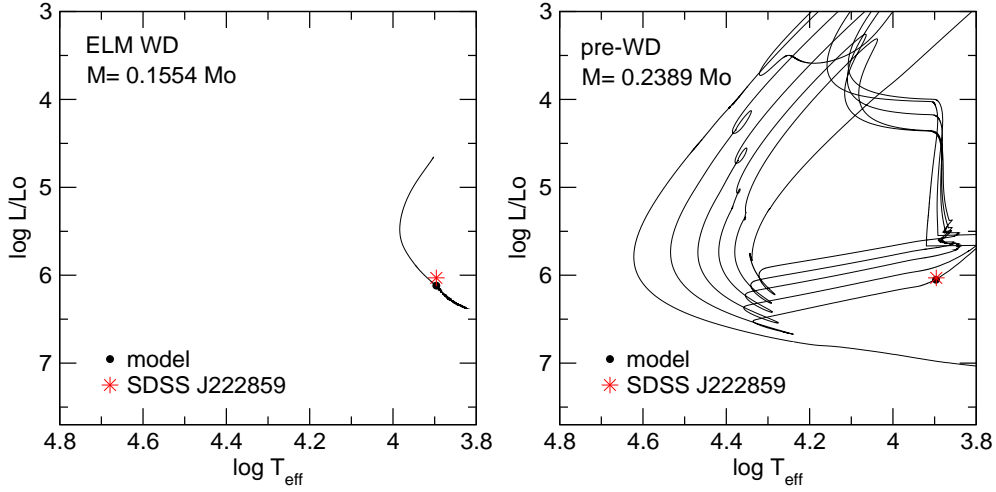} 
\caption{Location of the star SDSS J222859.93+362359.6 (red star symbol) 
in the $\log T_{\rm eff}-\log g$ diagram, along with the template ELM WD model
on its evolutionary track (left-hand panel) and its twin template pre-WD model
with virtually the same $T_{\rm eff}/ \log g$ values on its evolutionary 
track (right-hand panel).}
\label{figure-HR-J222} 
\end{center}
\end{figure*}

The pulsating ELM WD star SDSS J222859.93+362359.6 is the coolest
pulsating WD known so far,  with $T_{\rm eff}= 7870  \pm 120$ K. At
this low effective temperature, it is remarkably cooler than the other
pulsating objects of the class.   That this star  pulsates at
all is
somewhat surprising because there are many ELM WDs that do not pulsate
in between of this star and SDSS J161431.28+191219.4,  the second-coldest pulsating object ($T_{\rm eff}= 8800 \pm 170$ K).  Figure \ref{figure-Tracks_Multi_II} reveals the very
interesting  fact that the star SDSS J222859.93+362359.6  has $T_{\rm
  eff}$ and $\log g$ values that are degenerate with those  of more
massive pre-WDs that are undergoing a CNO flash  episode. This
occurs for  pre-WD tracks of the sequences with $M_*=
0.3624 M_{\odot}$,  $M_*= 0.3605 M_{\odot}$, $M_*= 0.2707 M_{\odot}$,
$M_*= 0.2389 M_{\odot}$, $M_*= 0.2019 M_{\odot}$, and  $M_*= 0.1805
M_{\odot}$\footnote{A similar situation is found  for
  J111215.82+111745.0  and SDSS J184037.78+642312.3, but in these
  cases the  position of the stars matches only one or two pre-WD
  tracks.}.  This makes it natural to wonder whether the star is a 
genuine ELM WD star
with a stellar mass of $M_* \sim 0.16 M_{\odot}$, or   is a
more massive pre-WD star going through a CNO flash.  The hypothesis that
this star might be a pre-WD is interesting on its own and deserves to be 
explored, even taking into account that 
the evolution  of the pre-WDs is much faster than that of the ELM WDs,  
and therefore there are far fewer opportunities of observing it.
To find some clue,  we computed
the pulsation spectrum of $g$ modes for  a template model belonging to
the sequence with $M_*= 0.2389 M_{\odot}$  when it is looping through
one of its CNO flashes and is briefly  located at $T_{\rm eff}= 7870$
K, $\log g= 6.03$.  In Fig. \ref{figure-HR-J222} we depict the
evolutionary track  of the $0.2389 M_{\odot}$ sequence and the
location of this template   model on the $T_{\rm eff}-\log g$
diagram. We  also include in the analysis  a template ELM WD model
with  similar $T_{\rm eff}$ and $\log g$ values and $M_*= 0.1554
M_{\odot}$  (see left-hand panel of Fig. \ref{figure-HR-J222}) to
compare  its pulsation spectrum with that of the $0.2389 M_{\odot}$
pre-WD model.

\begin{figure*} 
\begin{center}
\includegraphics[clip,width=14 cm]{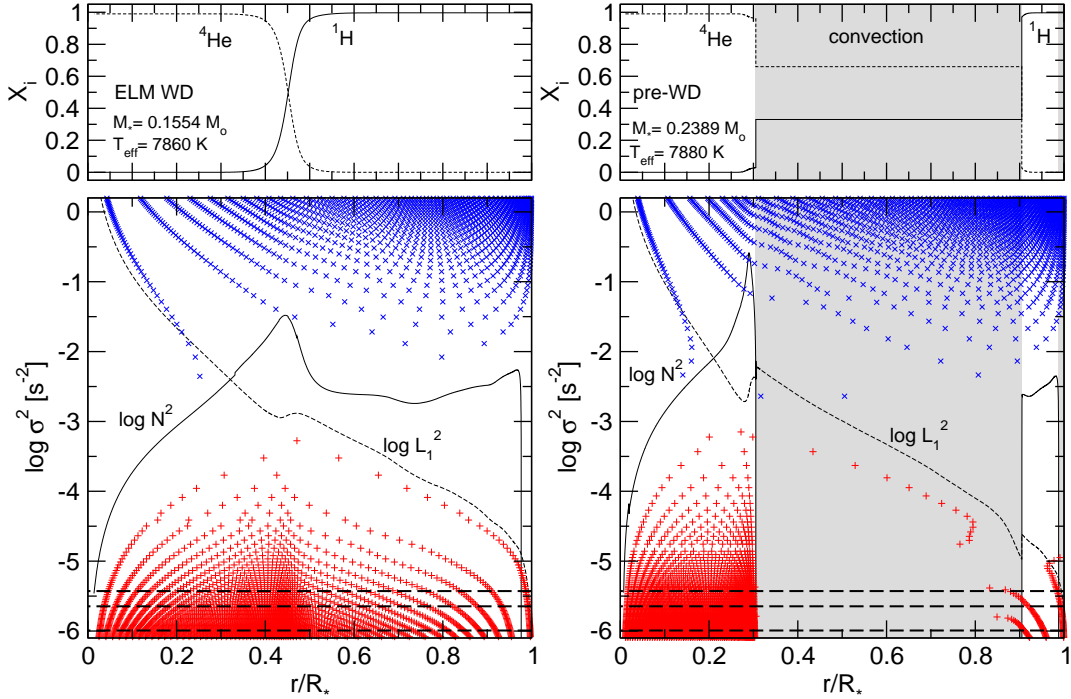} 
\caption{Internal chemical profiles of  He and H (upper panels) and
the propagation diagrams  (lower panels) 
for  the ELM WD template  model  with 
$M_*=  0.1554  M_{\odot}$ and  $T_{\rm eff}  \approx
7\,860$ K (left) and the pre-WD template model with 
$M_*=  0.2389  M_{\odot}$ and  a similar $T_{\rm eff}$ (right).
The grey areas in the $0.2389  M_{\odot}$ model indicate a thick inner convective zone at 
$0.3 \lesssim r/R_* \lesssim 0.9$ and another, very thin 
outer convection zone, at $0.988 \lesssim r/R_* \leq 1$. 
Dashed horizontal lines represent the three frequencies detected 
in SDSS J222859.93+362359.6.} 
\label{figure-xfbvl-j222859} 
\end{center}
\end{figure*}

In Fig. \ref{figure-xfbvl-j222859} we show the chemical profiles
and propagation diagrams of the two template models, 
including the three frequencies detected in SDSS J222859.93+362359.6. The strongly different internal 
chemical structure of both models is evident, as is expected because they represent 
very different evolutionary stages, although they share the same 
effective temperature and surface gravity. In particular, 
the pre-WD model (which is going through a
CNO flash episode) is characterised by a very thick internal
convective zone from $r/R_* \sim 0.3$ to $r/R_* \sim 0.9$.
This convection zone forces most $g$ modes to be strongly confined to the He core ($r/R_* \lesssim 0.3$), 
but a few $g$ modes are instead strongly trapped in the H envelope.
Note also the very thin outer convection zone, which
extends from $r/R_* \sim 0.988$ up to $r/R_*= 1$.  
Of course, $g$ modes are evanescent in the convective regions. 

\begin{figure*} 
\begin{center}
\includegraphics[clip,width=16 cm]{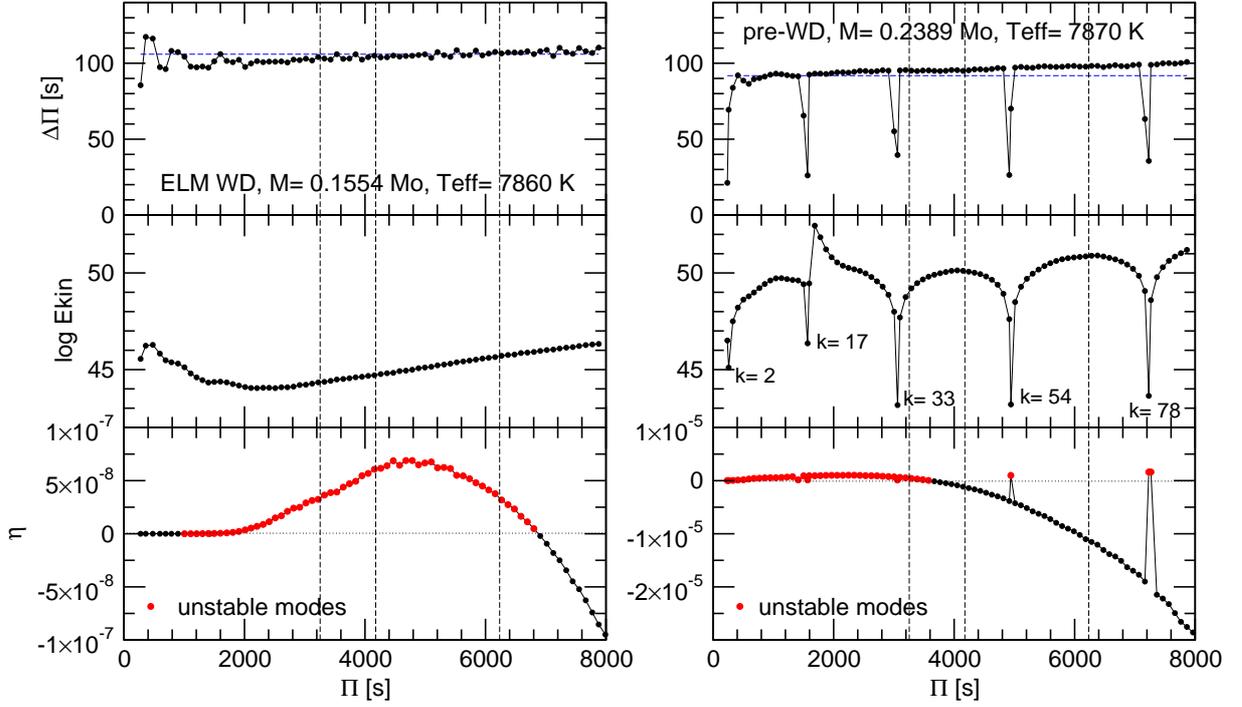} 
\caption{ $\ell= 1$ period spacing (upper panels), the 
kinetic oscillation energy (middle panels) and the growth rates   
(lower panels) in terms of the periods for the template ELM WD model (left) and the template 
pre-WD model (right). $\eta > 0$ implies pulsationally 
unstable modes. Vertical dashed lines represent the periods
observed in SDSS J222859.93+362359.6.}
\label{figure-delp-ekin-eta-j222859-gmodes} 
\end{center}
\end{figure*}

The period spacing and kinetic energy of the $g$ modes
with $\ell= 1$ in terms of periods for both template models 
are shown in the upper and middle panels of Fig. 
\ref{figure-delp-ekin-eta-j222859-gmodes}, in which we 
include, for illustrative purposes, the three periods observed in 
SDSS J222859.93+362359.6. In addition to a slight difference in the 
asymptotic period-spacing ($\Delta \Pi^{\rm a}= 106.5$ s for the 
ELM WD model, and $\Delta \Pi^{\rm a}= 91.85$ s for the 
pre-WD model),  
very pronnounced minima in $\Delta \Pi$ appear for the pre-WD
model, which are associated with modes with 
$k= 2$, $k= 17$, $k= 33$, $k= 54,$ and $k = 78$, which exhibit 
very low kinetic energies 
(up to six orders of magnitude lower than the average). These 
modes are strongly trapped in the outer (radiative) H envelope. They
should be easy to excite up to observable amplitudes by virtue 
of their very low kinetic oscillation energy (low inertia). 
If this were true, then the spectrum of the pre-WD model should 
be characterised by a few excited modes with clearly
separated periods instead of a continuous spectrum of unstable 
periods - which should be found,
instead, in the $0.1554 M_{\odot}$ ELM WD model, by extrapolating 
to the results of \citet{2012A&A...547A..96C} for an ELM WD model
with $M_*= 0.17 M_{\odot}$. If all this were true, the period spectrum of
SDSS J222859.93+362359.6, which is itself discrete and composed of only three periods, should be more nearly compatible with the pre-WD model 
than with the ELM WD model.  

\begin{figure} 
\begin{center}
\includegraphics[clip,width=9 cm]{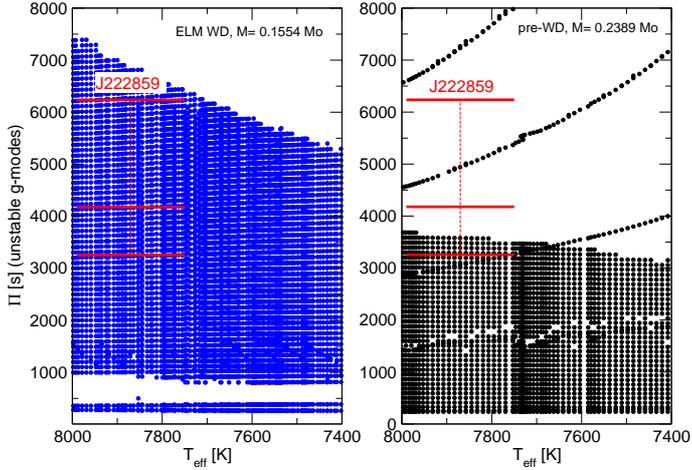} 
\caption{Instability domain  on the $T_{\rm eff} -
\Pi$ plane for $\ell = 1$ $g$ modes for the set of ELM WD 
models with $M_* = 0.1554 M_{\odot}$ (left) and the sequence of 
pre-WD models with $M_* = 0.2389 M_{\odot}$ (right). For the 
$M_* = 0.1554 M_{\odot}$ sequence, the modes with $k= 1$ and $k= 2$ are 
marginally 
unstable $\left(\eta \lesssim 10^{-10}\right)$. 
We also show the periodicities 
measured in SDSS J222859.93+362359.6, marked with red horizontal segments.}
\label{figure-p-teff} 
\end{center}
\end{figure}

To test this hypothesis, we computed non-adiabatic 
pulsations. Specifically, we performed an exploratory 
stability analysis of the ELM WD sequence with $M_*= 0.1554 M_{\odot}$ 
and the pre-WD sequence with $M_*= 0.2389 M_{\odot}$. These computations 
were carried out with the help of the non-adiabatic version of the 
{\tt LP-PUL} pulsation  code described in Sect. \ref{pulsation-codes}. 
We only computed $\ell= 1$ $g$ modes with 
periods in the range of interest and effective temperatures 
in the range  $9000 \gtrsim T_{\rm eff} \gtrsim  7400$ K, thus
 embracing the effective temperature of 
SDSS J222859.93+362359.6 ($T_{\rm eff} \sim 7900$ K).  
The output of our code are the normalised 
non-adiabatic growth rates, $\eta$, which are defined as 
$\eta = -\Im(\sigma)/\Re(\sigma)$, where $\Re(\sigma)$ 
and $\Im(\sigma)$ are the real
and the imaginary part of the complex eigenfrequency
$\sigma$ \citep[see][for details]{2006A&A...458..259C}. 
Positive values of $\eta$  mean unstable modes.
The growth rates 
in terms of the pulsation periods are shown
in the lower panels of Fig. \ref{figure-delp-ekin-eta-j222859-gmodes}.
For the $0.2389 M_{\odot}$ pre-WD model, the few modes
that are strongly trapped in the H envelope are unstable, as 
expected. In particular, modes with $k= 54$ and $k= 78$ are the only 
unstable modes for periods $\Pi \gtrsim 3600$ s. 
The differential work function 
\citep[see][for a definition]{2006A&A...458..259C} 
shows that most of driving for these modes comes
from  the base of the outer convective zone.
Below $\Pi \sim 3600$ s we found a continuous band of unstable 
modes with consecutive radial orders. The differential work 
functions for these 
unstable modes indicate that they are excited at the base of the
inner convective zone, except for those with $k= 2$, $k= 17$, and $k= 33$, 
which are trapped in the outer H envelope and are driven at the base
of the outer convection zone. On the other hand, the spectrum of unstable 
modes for the $0.1554 M_{\odot}$ ELM WD model consists
of  a continuous range of periods in the range 
$800 \lesssim \Pi \lesssim 7000$ s with consecutive radial
orders ($8 \leq k \leq 64$), much in line with the results of 
\citet{2012A&A...547A..96C}. The $k= 1$ and $k= 2$ modes 
are marginally unstable $\left(\eta \lesssim 10^{-10}\right)$. 
We note that the growth rates $\eta$ 
for the unstable modes of the $0.2389 M_{\odot}$ model 
are, on average, more than ten times larger than for the $0.1554 M_{\odot}$ 
model. This indicates that the unstable modes in the pre-WD model are more 
strongly excited than in the ELM WD model.
Finally, in Fig. \ref{figure-p-teff} a more general picture of the
situation is shwon. In this plot we depict the unstable periods  in
terms of the effective temperature for the sequence of pre-WD models
with $M_*= 0.1554 M_{\odot}$  and for the sequence of ELM WD
models with $M_*= 0.2389 M_{\odot}$ . We included the three
periods  detected in  SDSS J222859.93+362359.6. The discrete period spectrum of SDSS
J222859.93+362359.6 is  better represented \emph{qualitatively}  by the
period spectrum of unstable modes of   the $0.2389 M_{\odot}$ pre-WD
model sequence than by those  of  the  $0.1554 M_{\odot}$ ELM
WD sequence.

In summary, if we accept as valid the possibility that  SDSS
J222859.93+362359.6 might be a pre-WD star that we are only 
observing by chance at its brief incursion in the domain
of the $\log T_{\rm eff}-\log g$ diagram  where 
most ELM WDs are found, we would need to explain 
why  we do not observe the
continuous period spectrum as predicted by the non-adiabatic
computation for the range of periods  $200 \lesssim \Pi \lesssim 3600$
s.  Similarly, if we adopt the idea that SDSS J222859.93+362359.6 is a
genuine  ELM WD star, we need to explain why of the entire continuum
spectrum  of unstable periods, as predicted by theoretical models, we
only observe  three isolated periods. All in all, currently we
can only state that this star might be\emph{} a pre-WD instead of an ELM WD,
but more observations aimed at detecting additional periods, together with
extensive non-adiabatic calculations, are necessary  to
confirm or discard this statement.

\section{Summary and conclusions}
\label{conclusions}

We have  presented a  comprehensive theoretical
study of the seismic properties of low-mass, He-core WDs with
masses   in    the   range   $0.1554 - 0.4389   M_{\odot}$.  
We employed state-of-the-art evolutionary  stellar   structures  
representative  of   these  stars, extracted  from the  
sequences  of low-mass  He-core  WDs of  
\citet{2013A&A...557A..19A}. These  models were  
derived  by computing  the non-conservative
evolution of a binary  system consisting of an initially $1
M_{\odot}$ ZAMS  star and  a $1.4 M_{\odot}$  neutron star  
for various initial orbital periods. The evolutionary
computations were carried out by accounting  for a time-dependent
treatment  of the  gravitational settling  and chemical diffusion, as
well as of  residual nuclear burning.  We explored the adiabatic
pulsation properties of these  models, including the asymptotic 
predictions, the expected range of
periods, period spacings and frequency spacings,  the  propagation  
properties and  mode trapping of  pulsations, 
as  well as the dependence on the effective temperature and 
stellar mass and the effects of element diffusion.  
In particular, we  strongly emphasised the expected  differences in  the 
seismic  properties of objects with  $M_* \gtrsim  0.18 M_{\odot}$  
with CNO flashes during
the early-cooling phase, and  the ELM WDs
($M_* \lesssim 0.18 M_{\odot}$) without H flashes.  

The pulsation properties of low-mass He-core WDs have  been  explored  in 
detail recently  by \citet{2012A&A...547A..96C} on the basis of  a
set of evolutionary models derived  by  \citet{2009A&A...502..207A}  
considering  progenitor stars with 
supersolar metallicities and single-star  evolution. 
In the present work,  we recovered much of the results 
of \citet{2012A&A...547A..96C} and extended that study in  two ways.   
First,  we  explored not  only  the non-radial  $g$ mode
pulsation  spectrum of  low-mass WD  models, but  also considered non-radial  $p$ modes ($\ell= 1, 2$) 
and  radial  ($\ell= 0$)  pulsation modes.
Second, and thanks to the availability  of five WD model sequences with
progenitors  without   CNO-flashes  (see  Table
\ref{table1}),  here  we were  able  to  explore  in  detail  
the  pulsation properties of ELM  WDs, which are characterised by very 
thick  H envelopes. This is at variance with the work of 
\citet{2012A&A...547A..96C}, in which only  one WD  sequence  
(that  with mass  $M_*=  0.17 M_{\odot}$)  belonged to 
a progenitor star without H flashes. Finally, our study 
relied on evolutionary models consistent with the expected binary 
evolution of progenitor stars.
 
We also discussed  how our models match the  observed properties of
the known five pulsating low-mass  WD stars. In particular,
we tried to determine whether our models are able to account for 
the short periods observed in the star SDSS  J111215.82+111745.0, and  
evaluated  the possibility that these modes might be  $p$ modes and/ radial 
modes.  In addition, we tested the hypothesis
that one of  these stars, SDSS J222859.93+362359.6,  is  not a genuine
ELM  WD, but  instead  a pre-WD  star going  through  a CNO  flash
episode. 

Although we included ​​some exploratory non-adiabatic 
pulsation computations, most of our results rely on adiabatic 
pulsations. We defer a thorough
non-adiabatic exploration of our complete set of He-core WD models
to a forthcoming paper.

We summarise our findings below.
\begin{itemize}

\item Low-mass WDs have not been clearly classified as ELM WDs in the literature up to now. Here, we proposed to define as
ELM WDs the low-mass WDs without CNO
flashes in their early-cooling branch. According to this classification,
and in the frame of our computations, ELM WDs are the low-mass 
WDs with masses below $\sim 0.18 M_{\odot}$. We call stars with $M_* \gtrsim 0.18 M_{\odot}$ LM (low-mass) WDs. 

\item The asymptotic period-spacing of $g$ modes in low-mass He-core WDs
is larger for  lower mass and/or effective temperature.
The strong dependence of the period spacing on $M_*$ might be  used
to infer  the stellar mass  of  pulsating  low-mass   
WDs,  provided  that  enough consecutive pulsation periods of $g$ modes 
were detected, although this prospect is complicated by the  
fact that the period
spacing also depends on the thickness of the outer H envelope. 
In particular, there exist an ambiguity of $\Delta
\Pi_{\ell}^{\rm a}$ for masses near the threshold mass, 
$M_* \sim 0.18 M_{\odot}$.

\item The asymptotic frequency-spacing of $p$ modes and radial modes 
is larger for higher stellar masses  and lower effective temperatures.  
$\Delta \nu_{\ell}^{\rm a}$ is insensitive to the thickness of the H envelope.
If the detection of this type of modes were confirmed in future 
observations, the eventual measurement of the mean frequency-spacing 
for a real star might help to constrain  its stellar mass.

\item $g$ modes in ELM WDs ($M_* \lesssim 0.18 M_{\odot}$) 
mainly probe the core regions and $p$ modes the envelope, which provides the opportunity of constraining both the core and envelope chemical 
structure of these stars via asteroseismology. 

\item For LM WDs, $g$ modes are very sensitive to the He/H compositional 
gradient and therefore they can be a diagnostic tool for constraining the H-envelope thickness of low-mass WD models  with $M_* \gtrsim 0.18 M_{\odot}$.

\item The $g$ -mode periods are longer for smaller mass and lower effective 
temperature, and in the case of $p$  and radial modes, the periods 
increase with lower masses and higher effective temperatures.
In both cases, the dependence on the effective temperature
is much weaker than that on stellar mass.

\item Time-dependent element  diffusion strongly  affects the
$g$-mode pulsation spectrum of  low-mass WDs, in particular
ELM  WDs.  Diffusion processes substantially alter the shape of the 
He/H chemical interface and in turn the resulting  period spectrum.
The effects are weaker but still non-negligible for LM WDs. 
We claim that time-dependent element diffusion must  be  taken  into  
account  in any pulsational analysis of low-mass WD stars.

\item The chemical structure, propagation diagrams, and consequently
the pulsation properties of LM WD and ELM WD models with masses
near the limit mass ($M_∗ \sim 0.18 M_{\odot}$) are markedly different.
These differences are reflected in the period spacing of $g$
modes
and the frequency spacing of $p$ modes. For period
spacing of $g$ modes, it might be possible to use these differences 
as a practical tool to distinguish stars with CNO flashes
in their early-cooling phase from those without them, provided
that enough consecutive low- and intermediate-order $g$ modes
with the same harmonic degree were observed in pulsating low-mass
WDs. This optimistic view is dampened, however, because the most useful low radial-order $g$ modes are most likely 
pulsationally stable, as shown from the non-adiabatic analysis of
\citet{2012A&A...547A..96C}, and also demonstrated by 
most of the observed pulsating
low-mass WD stars, in which no periods shorter than about
1400 s are seen. Regarding the frequency spacing of $p$ modes,
$\Delta \nu^{\rm a}$ for the LM WD model is about
twice that of the ELM WD model, even though the difference
in stellar mass is only $\sim 0.004M_{\odot}$. This property might
constitute a useful seismological tool if many $p$ mode consecutive periods
in pulsating low-mass WDs were detected. 

\item Although there are only five pulsating low-mass WDs known at present,
and they are not enough in number to trace
clear trends of their pulsation spectra, we can note that
the observed periods are in general longer for lower gravities and
effective temperatures. These trends are in line with theoretical
considerations, as discussed by \citet{2013MNRAS.436.3573H}.

\item The star SDSS J111215.82+111745.0 exhibits short period
pulsations in the range $\sim 107-140$ s. For the temperature and mass 
(gravity) of this star according to spectroscopy, our models are unable to explain these short periods. In particular, 
if  the temperature and mass (gravity) of the star are correct, 
the short periods cannot be attributed to 
$p$ modes and/or radial modes. However, these periods might be 
caused by low-order $p$ modes and/or radial modes if the stellar 
mass were lower ($M_* \approx 0.16 M_{\odot}$). Alternatively, they might be 
low-order $g$ modes if the stellar 
mass were substantially larger ($\approx 0.43 M_{\odot}$). 
Another (unlikely) possibility is that at the spectroscopic mass (gravity) and 
$T_{\rm eff}$, the observed periods might be caused by 
higher-degree ($\ell \geq  3$) low-order $g$ modes.

\item The pulsating ELM WD SDSS J222859.93+362359.6 is the coolest
pulsating WD known so far. According to its location 
in the $\log T_{\rm eff}- \log g$ plane, this star might be a pre-WD star  that is looping through
one of its CNO flashes before it enters its final WD cooling track.
According to our models, we can neither confirm nor discard 
this hypothesis at present, although some indications of its discrete period
spectrum might promote the idea that the star is a 
pre-WD star. 

\end{itemize}

Pulsating low-mass white dwarfs have just begun to be discovered. 
Probably, more pulsating stars of this type will be 
detected soon, which will render them a very attractive target for asteroseismological 
studies. Asteroseismology of these stars will provide valuable 
clues about their internal structure and evolutionary status, 
allowing us to place constraints on the binary evolutionary 
processes involved in their formation. 
Needless to say, detailed evolutionary and
pulsational models like those presented here will be required 
to achieve this challenging goal.

\begin{acknowledgements}
We wish to thank our anonymous referee for the constructive
comments and suggestions that greatly improved the original version of
the paper. We also warmly thank J. J. Hermes for reading the paper and making
enlightening comments and suggestions.  Part of this work was
supported by AGENCIA through the Programa de Modernizaci\'on
Tecnol\'gica BID 1728/OC-AR, and by the PIP 112-200801-00940 grant
from CONICET. This research has made use of NASA Astrophysics Data
System.
\end{acknowledgements}

% for the bibliography, at the end

\bibliographystyle{aa} % style aa.bst
\bibliography{paper} % your references Yourfile.bib

\begin{thebibliography}{84}
\expandafter\ifx\csname natexlab\endcsname\relax\def\natexlab#1{#1}\fi

\bibitem[{{Althaus} \& {C{\'o}rsico}(2004)}]{2004A&A...417.1115A}
{Althaus}, L.~G. \& {C{\'o}rsico}, A.~H. 2004, \aap, 417, 1115

\bibitem[{{Althaus} {et~al.}(2004){Althaus}, {C{\'o}rsico}, {Gautschy}, {Han},
  {Serenelli}, \& {Panei}}]{2004MNRAS.347..125A}
{Althaus}, L.~G., {C{\'o}rsico}, A.~H., {Gautschy}, A., {et~al.} 2004, \mnras,
  347, 125

\bibitem[{{Althaus} {et~al.}(2010){Althaus}, {C{\'o}rsico}, {Isern}, \&
  {Garc{\'{\i}}a-Berro}}]{review}
{Althaus}, L.~G., {C{\'o}rsico}, A.~H., {Isern}, J., \& {Garc{\'{\i}}a-Berro},
  E. 2010, \aapr, 18, 471

\bibitem[{{Althaus} {et~al.}(2013){Althaus}, {Miller Bertolami}, \&
  {C{\'o}rsico}}]{2013A&A...557A..19A}
{Althaus}, L.~G., {Miller Bertolami}, M.~M., \& {C{\'o}rsico}, A.~H. 2013,
  \aap, 557, A19

\bibitem[{{Althaus} {et~al.}(2009){Althaus}, {Panei}, {Romero}, {Rohrmann},
  {C{\'o}rsico}, {Garc{\'{\i}}a-Berro}, \& {Miller
  Bertolami}}]{2009A&A...502..207A}
{Althaus}, L.~G., {Panei}, J.~A., {Romero}, A.~D., {et~al.} 2009, \aap, 502,
  207

\bibitem[{{Althaus} {et~al.}(2001){Althaus}, {Serenelli}, \&
  {Benvenuto}}]{2001MNRAS.323..471A}
{Althaus}, L.~G., {Serenelli}, A.~M., \& {Benvenuto}, O.~G. 2001, \mnras, 323,
  471

\bibitem[{{Althaus} {et~al.}(2005){Althaus}, {Serenelli}, {Panei},
  {C{\'o}rsico}, {Garc{\'{\i}}a-Berro}, \&
  {Sc{\'o}ccola}}]{2005A&A...435..631A}
{Althaus}, L.~G., {Serenelli}, A.~M., {Panei}, J.~A., {et~al.} 2005, \aap, 435,
  631

\bibitem[{{Bassa}(2006)}]{2006PhDT.........7B}
{Bassa}, C.~G. 2006, PhD thesis, Astronomical Institute, Utrecht University, PO
  Box 80 000, 3508 TA Utrecht, The Netherlands

\bibitem[{{Bassa} {et~al.}(2003){Bassa}, {van Kerkwijk}, \&
  {Kulkarni}}]{2003A&A...403.1067B}
{Bassa}, C.~G., {van Kerkwijk}, M.~H., \& {Kulkarni}, S.~R. 2003, \aap, 403,
  1067

\bibitem[{{Beauchamp} {et~al.}(1999){Beauchamp}, {Wesemael}, {Bergeron},
  {Fontaine}, {Saffer}, {Liebert}, \& {Brassard}}]{1999ApJ...516..887B}
{Beauchamp}, A., {Wesemael}, F., {Bergeron}, P., {et~al.} 1999, \apj, 516, 887

\bibitem[{{Bradley} {et~al.}(1993){Bradley}, {Winget}, \&
  {Wood}}]{1993ApJ...406..661B}
{Bradley}, P.~A., {Winget}, D.~E., \& {Wood}, M.~A. 1993, \apj, 406, 661

\bibitem[{{Brassard} \& {Fontaine}(1999)}]{1999ASPC..173..329B}
{Brassard}, P. \& {Fontaine}, G. 1999, in Astronomical Society of the Pacific
  Conference Series, Vol. 173, Stellar Structure: Theory and Test of Connective
  Energy Transport, ed. A.~{Gimenez}, E.~F. {Guinan}, \& B.~{Montesinos}, 329

\bibitem[{{Brassard} {et~al.}(1992){Brassard}, {Fontaine}, {Wesemael}, \&
  {Hansen}}]{1992ApJS...80..369B}
{Brassard}, P., {Fontaine}, G., {Wesemael}, F., \& {Hansen}, C.~J. 1992, \apjs,
  80, 369

\bibitem[{{Brickhill}(1990)}]{1990MNRAS.246..510B}
{Brickhill}, A.~J. 1990, \mnras, 246, 510

\bibitem[{{Brown} {et~al.}(2013){Brown}, {Kilic}, {Allende Prieto},
  {Gianninas}, \& {Kenyon}}]{2013ApJ...769...66B}
{Brown}, W.~R., {Kilic}, M., {Allende Prieto}, C., {Gianninas}, A., \&
  {Kenyon}, S.~J. 2013, \apj, 769, 66

\bibitem[{{Brown} {et~al.}(2010){Brown}, {Kilic}, {Allende Prieto}, \&
  {Kenyon}}]{2010ApJ...723.1072B}
{Brown}, W.~R., {Kilic}, M., {Allende Prieto}, C., \& {Kenyon}, S.~J. 2010,
  \apj, 723, 1072

\bibitem[{{Brown} {et~al.}(2012){Brown}, {Kilic}, {Allende Prieto}, \&
  {Kenyon}}]{2012ApJ...744..142B}
---. 2012, \apj, 744, 142

\bibitem[{{Burgers}(1969)}]{1969fecg.book.....B}
{Burgers}, J.~M. 1969, {Flow Equations for Composite Gases} (New York: Academic
  Press)

\bibitem[{{Cassisi} {et~al.}(2007){Cassisi}, {Potekhin}, {Pietrinferni},
  {Catelan}, \& {Salaris}}]{2007ApJ...661.1094C}
{Cassisi}, S., {Potekhin}, A.~Y., {Pietrinferni}, A., {Catelan}, M., \&
  {Salaris}, M. 2007, \apj, 661, 1094

\bibitem[{{Chang} {et~al.}(2013){Chang}, {Shih}, {Liu}, {Fan}, {Wu}, {Roques},
  {Doressoundiram}, {Fernandez}, {Christophe}, \&
  {Dauny}}]{2013A&A...558A..63C}
{Chang}, H.-K., {Shih}, I.-C., {Liu}, C.-Y., {et~al.} 2013, \aap, 558, A63

\bibitem[{{Chen} \& {Han}(2002)}]{2002MNRAS.335..948C}
{Chen}, X. \& {Han}, Z. 2002, \mnras, 335, 948

\bibitem[{{Cohen} {et~al.}(1969){Cohen}, {Lapidus}, \&
  {Cameron}}]{1969Ap&SS...5..113C}
{Cohen}, J.~M., {Lapidus}, A., \& {Cameron}, A.~G.~W. 1969, \apss, 5, 113

\bibitem[{{C{\'o}rsico} \& {Althaus}(2006)}]{2006A&A...454..863C}
{C{\'o}rsico}, A.~H. \& {Althaus}, L.~G. 2006, \aap, 454, 863

\bibitem[{{C{\'o}rsico} {et~al.}(2002{\natexlab{a}}){C{\'o}rsico}, {Althaus},
  {Benvenuto}, \& {Serenelli}}]{2002A&A...387..531C}
{C{\'o}rsico}, A.~H., {Althaus}, L.~G., {Benvenuto}, O.~G., \& {Serenelli},
  A.~M. 2002{\natexlab{a}}, \aap, 387, 531

\bibitem[{{C{\'o}rsico} {et~al.}(2006){C{\'o}rsico}, {Althaus}, \& {Miller
  Bertolami}}]{2006A&A...458..259C}
{C{\'o}rsico}, A.~H., {Althaus}, L.~G., \& {Miller Bertolami}, M.~M. 2006,
  \aap, 458, 259

\bibitem[{{C{\'o}rsico} {et~al.}(2009{\natexlab{a}}){C{\'o}rsico}, {Althaus},
  {Miller Bertolami}, \& {Garc{\'{\i}}a-Berro}}]{2009JPhCS.172a2075C}
{C{\'o}rsico}, A.~H., {Althaus}, L.~G., {Miller Bertolami}, M.~M., \&
  {Garc{\'{\i}}a-Berro}, E. 2009{\natexlab{a}}, Journal of Physics Conference
  Series, 172, 012075

\bibitem[{{C{\'o}rsico} {et~al.}(2002{\natexlab{b}}){C{\'o}rsico}, {Benvenuto},
  {Althaus}, \& {Serenelli}}]{2002MNRAS.332..392C}
{C{\'o}rsico}, A.~H., {Benvenuto}, O.~G., {Althaus}, L.~G., \& {Serenelli},
  A.~M. 2002{\natexlab{b}}, \mnras, 332, 392

\bibitem[{{C{\'o}rsico} {et~al.}(2009{\natexlab{b}}){C{\'o}rsico}, {Romero},
  {Althaus}, \& {Garc{\'{\i}}a-Berro}}]{2009A&A...506..835C}
{C{\'o}rsico}, A.~H., {Romero}, A.~D., {Althaus}, L.~G., \&
  {Garc{\'{\i}}a-Berro}, E. 2009{\natexlab{b}}, \aap, 506, 835

\bibitem[{{C{\'o}rsico} {et~al.}(2012){C{\'o}rsico}, {Romero}, {Althaus}, \&
  {Hermes}}]{2012A&A...547A..96C}
{C{\'o}rsico}, A.~H., {Romero}, A.~D., {Althaus}, L.~G., \& {Hermes}, J.~J.
  2012, \aap, 547, A96

\bibitem[{{Cox}(1980)}]{1980tsp..book.....C}
{Cox}, J.~P. 1980, {Theory of stellar pulsation}

\bibitem[{{Driebe} {et~al.}(1998){Driebe}, {Schoenberner}, {Bloecker}, \&
  {Herwig}}]{1998A&A...339..123D}
{Driebe}, T., {Schoenberner}, D., {Bloecker}, T., \& {Herwig}, F. 1998, \aap,
  339, 123

\bibitem[{{Dziembowski}(1977)}]{1977AcA....27..203D}
{Dziembowski}, W. 1977, \actaa, 27, 203

\bibitem[{{Dziembowski}(1971)}]{1971AcA....21..289D}
{Dziembowski}, W.~A. 1971, \actaa, 21, 289

\bibitem[{{Ferguson} {et~al.}(2005){Ferguson}, {Alexander}, {Allard}, {Barman},
  {Bodnarik}, {Hauschildt}, {Heffner-Wong}, \& {Tamanai}}]{2005ApJ...623..585F}
{Ferguson}, J.~W., {Alexander}, D.~R., {Allard}, F., {et~al.} 2005, \apj, 623,
  585

\bibitem[{{Fontaine} \& {Brassard}(2008)}]{2008PASP..120.1043F}
{Fontaine}, G. \& {Brassard}, P. 2008, \pasp, 120, 1043

\bibitem[{{Garc{\'{\i}}a-Berro} {et~al.}(2010){Garc{\'{\i}}a-Berro}, {Torres},
  {Althaus}, {Renedo}, {Lor{\'e}n-Aguilar}, {C{\'o}rsico}, {Rohrmann},
  {Salaris}, \& {Isern}}]{nature}
{Garc{\'{\i}}a-Berro}, E., {Torres}, S., {Althaus}, L.~G., {et~al.} 2010, \nat,
  465, 194

\bibitem[{{Grevesse} \& {Sauval}(1998)}]{1998SSRv...85..161G}
{Grevesse}, N. \& {Sauval}, A.~J. 1998, \ssr, 85, 161

\bibitem[{{Haft} {et~al.}(1994){Haft}, {Raffelt}, \&
  {Weiss}}]{1994ApJ...425..222H}
{Haft}, M., {Raffelt}, G., \& {Weiss}, A. 1994, \apj, 425, 222

\bibitem[{{Hermes} {et~al.}(2013{\natexlab{a}}){Hermes}, {Montgomery},
  {Gianninas}, {Winget}, {Brown}, {Harrold}, {Bell}, {Kenyon}, {Kilic}, \&
  {Castanheira}}]{2013MNRAS.436.3573H}
{Hermes}, J.~J., {Montgomery}, M.~H., {Gianninas}, A., {et~al.}
  2013{\natexlab{a}}, \mnras, 436, 3573

\bibitem[{{Hermes} {et~al.}(2013{\natexlab{b}}){Hermes}, {Montgomery},
  {Winget}, {Brown}, {Gianninas}, {Kilic}, {Kenyon}, {Bell}, \&
  {Harrold}}]{2013ApJ...765..102H}
{Hermes}, J.~J., {Montgomery}, M.~H., {Winget}, D.~E., {et~al.}
  2013{\natexlab{b}}, \apj, 765, 102

\bibitem[{{Hermes} {et~al.}(2012){Hermes}, {Montgomery}, {Winget}, {Brown},
  {Kilic}, \& {Kenyon}}]{2012ApJ...750L..28H}
---. 2012, \apjl, 750, L28

\bibitem[{{Iglesias} \& {Rogers}(1996)}]{1996ApJ...464..943I}
{Iglesias}, C.~A. \& {Rogers}, F.~J. 1996, \apj, 464, 943

\bibitem[{{Itoh} {et~al.}(1996){Itoh}, {Hayashi}, {Nishikawa}, \&
  {Kohyama}}]{1996ApJS..102..411I}
{Itoh}, N., {Hayashi}, H., {Nishikawa}, A., \& {Kohyama}, Y. 1996, \apjs, 102,
  411

\bibitem[{{Kawaler}(1993)}]{1993ApJ...404..294K}
{Kawaler}, S.~D. 1993, \apj, 404, 294

\bibitem[{{Kawaler} {et~al.}(1994){Kawaler}, {Bond}, {Sherbert}, \&
  {Watson}}]{1994AJ....107..298K}
{Kawaler}, S.~D., {Bond}, H.~E., {Sherbert}, L.~E., \& {Watson}, T.~K. 1994,
  \aj, 107, 298

\bibitem[{{Kepler} {et~al.}(2007){Kepler}, {Kleinman}, {Nitta}, {Koester},
  {Castanheira}, {Giovannini}, {Costa}, \& {Althaus}}]{2007MNRAS.375.1315K}
{Kepler}, S.~O., {Kleinman}, S.~J., {Nitta}, A., {et~al.} 2007, \mnras, 375,
  1315

\bibitem[{{Kilic} {et~al.}(2011){Kilic}, {Brown}, {Allende Prieto},
  {Ag{\"u}eros}, {Heinke}, \& {Kenyon}}]{2011ApJ...727....3K}
{Kilic}, M., {Brown}, W.~R., {Allende Prieto}, C., {et~al.} 2011, \apj, 727, 3

\bibitem[{{Kilic} {et~al.}(2012){Kilic}, {Brown}, {Allende Prieto}, {Kenyon},
  {Heinke}, {Ag{\"u}eros}, \& {Kleinman}}]{2012ApJ...751..141K}
---. 2012, \apj, 751, 141

\bibitem[{{Kilkenny} {et~al.}(2014){Kilkenny}, {Welsh}, {Koen}, {Gulbis}, \&
  {Kotze}}]{2014MNRAS.437.1836K}
{Kilkenny}, D., {Welsh}, B.~Y., {Koen}, C., {Gulbis}, A.~A.~S., \& {Kotze},
  M.~M. 2014, \mnras, 437, 1836

\bibitem[{{Kippenhahn} {et~al.}(2013){Kippenhahn}, {Weigert}, \&
  {Weiss}}]{2013sse..book.....K}
{Kippenhahn}, R., {Weigert}, A., \& {Weiss}, A. 2013, {Stellar Structure and
  Evolution} (Springer-Verlag Berlin Heidelberg)

\bibitem[{{Kleinman} {et~al.}(2013){Kleinman}, {Kepler}, {Koester}, {Pelisoli},
  {Pe{\c c}anha}, {Nitta}, {Costa}, {Krzesinski}, {Dufour}, {Lachapelle},
  {Bergeron}, {Yip}, {Harris}, {Eisenstein}, {Althaus}, \&
  {C{\'o}rsico}}]{2013ApJS..204....5K}
{Kleinman}, S.~J., {Kepler}, S.~O., {Koester}, D., {et~al.} 2013, \apjs, 204, 5

\bibitem[{{Koester} {et~al.}(2009){Koester}, {Voss}, {Napiwotzki},
  {Christlieb}, {Homeier}, {Lisker}, {Reimers}, \&
  {Heber}}]{2009A&A...505..441K}
{Koester}, D., {Voss}, B., {Napiwotzki}, R., {et~al.} 2009, \aap, 505, 441

\bibitem[{{Landolt}(1968)}]{1968ApJ...153..151L}
{Landolt}, A.~U. 1968, \apj, 153, 151

\bibitem[{{Ledoux} \& {Sauvenier-Goffin}(1950)}]{1950ApJ...111..611L}
{Ledoux}, P.~J. \& {Sauvenier-Goffin}, E. 1950, \apj, 111, 611

\bibitem[{{Magni} \& {Mazzitelli}(1979)}]{1979A&A....72..134M}
{Magni}, G. \& {Mazzitelli}, I. 1979, \aap, 72, 134

\bibitem[{{Marsh} {et~al.}(1995){Marsh}, {Dhillon}, \&
  {Duck}}]{1995MNRAS.275..828M}
{Marsh}, T.~R., {Dhillon}, V.~S., \& {Duck}, S.~R. 1995, \mnras, 275, 828

\bibitem[{{Maxted} {et~al.}(2011){Maxted}, {Anderson}, {Burleigh}, {Collier
  Cameron}, {Heber}, {G{\"a}nsicke}, {Geier}, {Kupfer}, {Marsh}, {Nelemans},
  {O'Toole}, {{\O}stensen}, {Smalley}, \& {West}}]{2011MNRAS.418.1156M}
{Maxted}, P.~F.~L., {Anderson}, D.~R., {Burleigh}, M.~R., {et~al.} 2011,
  \mnras, 418, 1156

\bibitem[{{Montgomery} {et~al.}(2008){Montgomery}, {Williams}, {Winget},
  {Dufour}, {De Gennaro}, \& {Liebert}}]{2008ApJ...678L..51M}
{Montgomery}, M.~H., {Williams}, K.~A., {Winget}, D.~E., {et~al.} 2008, \apjl,
  678, L51

\bibitem[{{Mukadam} {et~al.}(2006){Mukadam}, {Montgomery}, {Winget}, {Kepler},
  \& {Clemens}}]{2006ApJ...640..956M}
{Mukadam}, A.~S., {Montgomery}, M.~H., {Winget}, D.~E., {Kepler}, S.~O., \&
  {Clemens}, J.~C. 2006, \apj, 640, 956

\bibitem[{{Muslimov} \& {Sarna}(1993)}]{1993MNRAS.262..164M}
{Muslimov}, A.~G. \& {Sarna}, M.~J. 1993, \mnras, 262, 164

\bibitem[{{Ostriker} \& {Tassoul}(1969)}]{1969ApJ...155..987O}
{Ostriker}, J.~P. \& {Tassoul}, J.~L. 1969, \apj, 155, 987

\bibitem[{{Panei} {et~al.}(2007){Panei}, {Althaus}, {Chen}, \&
  {Han}}]{2007MNRAS.382..779P}
{Panei}, J.~A., {Althaus}, L.~G., {Chen}, X., \& {Han}, Z. 2007, \mnras, 382,
  779

\bibitem[{{Robinson}(1984)}]{1984AJ.....89.1732R}
{Robinson}, E.~L. 1984, \aj, 89, 1732

\bibitem[{{Rohrmann} {et~al.}(2012){Rohrmann}, {Althaus},
  {Garc{\'{\i}}a-Berro}, {C{\'o}rsico}, \& {Miller
  Bertolami}}]{2012A&A...546A.119R}
{Rohrmann}, R.~D., {Althaus}, L.~G., {Garc{\'{\i}}a-Berro}, E., {C{\'o}rsico},
  A.~H., \& {Miller Bertolami}, M.~M. 2012, \aap, 546, A119

\bibitem[{{Romero} {et~al.}(2012){Romero}, {C{\'o}rsico}, {Althaus}, {Kepler},
  {Castanheira}, \& {Miller Bertolami}}]{2012MNRAS.420.1462R}
{Romero}, A.~D., {C{\'o}rsico}, A.~H., {Althaus}, L.~G., {et~al.} 2012, \mnras,
  420, 1462

\bibitem[{{Romero} {et~al.}(2013){Romero}, {Kepler}, {C{\'o}rsico}, {Althaus},
  \& {Fraga}}]{2013ApJ...779...58R}
{Romero}, A.~D., {Kepler}, S.~O., {C{\'o}rsico}, A.~H., {Althaus}, L.~G., \&
  {Fraga}, L. 2013, \apj, 779, 58

\bibitem[{{Saio} {et~al.}(1983){Saio}, {Winget}, \&
  {Robinson}}]{1983ApJ...265..982S}
{Saio}, H., {Winget}, D.~E., \& {Robinson}, E.~L. 1983, \apj, 265, 982

\bibitem[{{Sarna} {et~al.}(2000){Sarna}, {Ergma}, \& {Ger{\v s}kevit{\v
  s}-Antipova}}]{2000MNRAS.316...84S}
{Sarna}, M.~J., {Ergma}, E., \& {Ger{\v s}kevit{\v s}-Antipova}, J. 2000,
  \mnras, 316, 84

\bibitem[{{Siess}(2007)}]{2007A&A...476..893S}
{Siess}, L. 2007, \aap, 476, 893

\bibitem[{{Silvotti} {et~al.}(2011){Silvotti}, {Fontaine}, {Pavlov}, {Marsh},
  {Dhillon}, {Littlefair}, \& {Getman}}]{2011A&A...525A..64S}
{Silvotti}, R., {Fontaine}, G., {Pavlov}, M., {et~al.} 2011, \aap, 525, A64

\bibitem[{{Silvotti} {et~al.}(2012){Silvotti}, {{\O}stensen}, {Bloemen},
  {Telting}, {Heber}, {Oreiro}, {Reed}, {Farris}, {O'Toole}, {Lanteri},
  {Degroote}, {Hu}, {Baran}, {Hermes}, {Althaus}, {Marsh}, {Charpinet}, {Li},
  {Morris}, \& {Sanderfer}}]{2012MNRAS.424.1752S}
{Silvotti}, R., {{\O}stensen}, R.~H., {Bloemen}, S., {et~al.} 2012, \mnras,
  424, 1752

\bibitem[{{Starrfield} {et~al.}(1983){Starrfield}, {Cox}, {Hodson}, \&
  {Clancy}}]{1983ApJ...269..645S}
{Starrfield}, S., {Cox}, A.~N., {Hodson}, S.~W., \& {Clancy}, S.~P. 1983, \apj,
  269, 645

\bibitem[{{Steinfadt} {et~al.}(2010){Steinfadt}, {Bildsten}, \&
  {Arras}}]{2010ApJ...718..441S}
{Steinfadt}, J.~D.~R., {Bildsten}, L., \& {Arras}, P. 2010, \apj, 718, 441

\bibitem[{{Steinfadt} {et~al.}(2012){Steinfadt}, {Bildsten}, {Kaplan},
  {Fulton}, {Howell}, {Marsh}, {Ofek}, \& {Shporer}}]{2012PASP..124....1S}
{Steinfadt}, J.~D.~R., {Bildsten}, L., {Kaplan}, D.~L., {et~al.} 2012, \pasp,
  124, 1

\bibitem[{{Tassoul} {et~al.}(1990){Tassoul}, {Fontaine}, \&
  {Winget}}]{1990ApJS...72..335T}
{Tassoul}, M., {Fontaine}, G., \& {Winget}, D.~E. 1990, \apjs, 72, 335

\bibitem[{{Tremblay} {et~al.}(2011){Tremblay}, {Bergeron}, \&
  {Gianninas}}]{2011ApJ...730..128T}
{Tremblay}, P.-E., {Bergeron}, P., \& {Gianninas}, A. 2011, \apj, 730, 128

\bibitem[{{Unno} {et~al.}(1989){Unno}, {Osaki}, {Ando}, {Saio}, \&
  {Shibahashi}}]{1989nos..book.....U}
{Unno}, W., {Osaki}, Y., {Ando}, H., {Saio}, H., \& {Shibahashi}, H. 1989,
  {Nonradial oscillations of stars}, ed. T.~University~of Tokyo~Press

\bibitem[{{Van Grootel} {et~al.}(2012){Van Grootel}, {Dupret}, {Fontaine},
  {Brassard}, {Grigahc{\`e}ne}, \& {Quirion}}]{2012A&A...539A..87V}
{Van Grootel}, V., {Dupret}, M.-A., {Fontaine}, G., {et~al.} 2012, \aap, 539,
  A87

\bibitem[{{Van Grootel} {et~al.}(2013){Van Grootel}, {Fontaine}, {Brassard}, \&
  {Dupret}}]{2013ApJ...762...57V}
{Van Grootel}, V., {Fontaine}, G., {Brassard}, P., \& {Dupret}, M.-A. 2013,
  \apj, 762, 57

\bibitem[{{van Kerkwijk} {et~al.}(2005){van Kerkwijk}, {Bassa}, {Jacoby}, \&
  {Jonker}}]{2005ASPC..328..357V}
{van Kerkwijk}, M.~H., {Bassa}, C.~G., {Jacoby}, B.~A., \& {Jonker}, P.~G.
  2005, in Astronomical Society of the Pacific Conference Series, Vol. 328,
  Binary Radio Pulsars, ed. F.~A. {Rasio} \& I.~H. {Stairs}, 357

\bibitem[{{Vauclair}(1971{\natexlab{a}})}]{1971IAUS...42..145V}
{Vauclair}, G. 1971{\natexlab{a}}, in IAU Symposium, Vol.~42, White Dwarfs, ed.
  W.~J. {Luyten}, 145

\bibitem[{{Vauclair}(1971{\natexlab{b}})}]{1971ApL.....9..161V}
{Vauclair}, G. 1971{\natexlab{b}}, \aplett, 9, 161

\bibitem[{{Weiss} \& {Ferguson}(2009)}]{2009A&A...508.1343W}
{Weiss}, A. \& {Ferguson}, J.~W. 2009, \aap, 508, 1343

\bibitem[{{Winget} \& {Kepler}(2008)}]{2008ARA&A..46..157W}
{Winget}, D.~E. \& {Kepler}, S.~O. 2008, \araa, 46, 157

\end{thebibliography}

\end{document}